\documentclass[a4paper,11pt]{article}
\pdfoutput=1 
\usepackage{amssymb,mathrsfs}
\usepackage{bbm,bm}
\usepackage{siunitx}
\usepackage{amsmath}
\usepackage{braket}
\usepackage{mathtools}
\usepackage[usenames,dvipsnames,svgnames,table]{xcolor}
\usepackage [utf8] {inputenc}
\usepackage{color}
\usepackage{lmodern}
\usepackage{footnote}
\usepackage{slashed}
\usepackage{mathrsfs}
\usepackage[pdftex] {graphicx}
\usepackage{multirow}
\usepackage{jheppub} 
\usepackage{here}
\usepackage{hyperref}
\usepackage{soul}
\usepackage[justification=centering, singlelinecheck=false]{caption}
\usepackage[list=true, labelfont=bf, labelformat=brace, position=top]{subcaption}
\newcommand{\eq}{\begin{eqnarray}}
\newcommand{\en}{\end{eqnarray}}

\title{Form factors of the $\bm{\rho}$ meson from effective field theory and the lattice}

\renewcommand{\thefootnote}{\fnsymbol{footnote}}

\author[1,2,3]{Ulf-G. Mei{\ss}ner}
\affiliation[1]{Helmholtz-Institut f\"ur Strahlen- und Kernphysik (Theorie)\\ and Bethe Center for Theoretical Physics, Universit\"at Bonn, 53115 Bonn, Germany}
\affiliation[2]{Institute for Advanced Simulation (IAS-4) Forschungszentrum Jülich, Germany}
\affiliation[3]{Peng Huanwu Collaborative Center for Research and Education,
International Institute for Interdisciplinary and Frontiers, Beihang University, Beijing 100191, China}
\emailAdd{meissner@hiskp.uni-bonn.de}

\author[1]{Akaki Rusetsky,}
\emailAdd{rusetsky@hiskp.uni-bonn.de}

\author[1]{Ajay S. Sakthivasan,}
\emailAdd{sakthivasan@hiskp.uni-bonn.de}

\author[4]{Gerrit Schierholz,}
\affiliation[4]{Deutsches Elektronen-Synchrotron DESY, Notkestr. 85, 22607 Hamburg, Germany.}
\emailAdd{gerrit.schierholz@desy.de}

\author[5]{and Jia-Jun Wu}
\affiliation[5]{School of Physical Sciences, University of Chinese Academy of Sciences, Beijing 100049, China}
\emailAdd{wujiajun@ucas.ac.cn}

\preprint{DESY-26-030}

\abstract{

\noindent
The calculation of resonance form factors in effective field theory as well as on the lattice is a highly challenging task. In a recent paper~\cite{Lozano:2022kfz}, we proposed a novel method based on the introduction of a background field and the Feynman-Hellmann theorem to address the problem, and applied it to a toy model. In the present work we use this method for the electromagnetic form factors of the $\rho$-meson. By matching the results to Chiral Perturbation Theory, we provide a first, crude estimate of all three form factors of the $\rho$-meson within the effective field theory. Contact contributions to these form factors turn out to be substantial. A procedure for lattice calculations is outlined, paving the way for an ab initio approach to the problem.

}

\allowdisplaybreaks
\begin{document}
\maketitle
\flushbottom

\renewcommand{\thefootnote}{\arabic{footnote}}

\section{Introduction}

Form factors provide valuable information on how quarks and gluons bind to hadrons.
This applies primarily to hadrons with higher spin, which are particularly informative.
Note that, apart from the lowest baryon and pseudoscalar octets, all hadrons of interest
are unstable in QCD. However, previous lattice calculations were largely limited to
the large quark masses~\cite{QCDSF:2008tjq,Alexandrou:2008bn,Stokes:2019zdd,Hall:2016kou}, which
rendered the resonances stable. For a review of unstable particles on the lattice, see~\cite{Mai:2022eur}.

In this paper we consider resonances in systems of two particles, which are themselves
stable under the strong interactions. Our main focus here is on the conceptual challenge
rather than the technical implementation. We namely propose to calculate the matrix
element of the electromagnetic current on the lattice by using the background field method (the Feynman-Hellmann theorem), which relates this matrix element to the shift of the discrete energy spectrum
in a finite box in the presence of an external current (this method is similar in spirit to the
L\"uscher formalism~\cite{Luscher:1990ux} that relates the infinite-volume
scattering phase shift in a two-particle system to the finite-volume energy spectrum.)
The relation between the infinite-volume current matrix element and the finite volume
spectrum, which serves as a basis for future lattice calculations, will be rigorously
derived here within a non-relativistic effective field theory (NREFT) framework.
Note that the Feynman-Hellmann theorem has proven to be a powerful tool in lattice and background field computations of the polarizabilities of stable particles~\cite{Bignell:2020dze,Primer:2013pva}, the magnetic properties of vector, axial, and tensor mesons~\cite{Lee:2008qf} and
of light nuclei~\cite{Chang:2015qxa}, as well as nucleon form factors, nucleon structure functions and generalized parton
distributions~\cite{QCDSF:2017ssq,Batelaan:2023jqp,Can:2025zsr,Can:2025jzf,CSSMQCDSFUKQCD:2021lkf}. For the resonances, the method was first tested in a non-relativistic toy model~\cite{Lozano:2022kfz}.

For our first physical application, we will focus on the electromagnetic form factors of the
charged $\rho$-meson, which practically is a two-pion resonance. It has three complex-valued form
factors, the linear combinations of which yield the electric, magnetic and quadrupole form factors. In this paper, we first present
the calculation of all three form factors in the infinite volume in NREFT. The crucial
point here is that we
were able to match the results with lowest order Chiral Perturbation Theory (ChPT), which
provides a quantitative estimate of all three form factors. Next, we present a
framework for calculating the resonance form factors on the lattice,
which will be needed when extracting the infinite-volume form factors out from the numerical simulations.

First, a few technical details are in order. In NREFT, any form factor splits naturally
into two clearly distinct contributions: the triangle diagram, in which the external
photon is attached to the charged pion, and the contact interaction (also referred to as short-range contribution in the literature), 
in which the photon couples to a local four-pion operator consisting of two charged and two neutral
pions (see, e.g.,~\cite{Lozano:2022kfz}). On the lattice, this corresponds to the photon
being coupled to one of the sea quarks.\footnote{Albeit, strictly speaking, there is no 
one-to-one correspondence between the diagrams in QCD and in NREFT. However, such an
interpretation provides a clear intuitive picture of various contributions to the form
factor.} The triangle diagram does not have a well-defined infinite volume
limit even if the Lellouch-L\"uscher factors for the external legs~\cite{Lellouch:2000pv} are properly taken into account\,\footnote{Note that this problem does not occur for transition form factors, where one of the particles is stable~\cite{Agadjanov:2014kha,Agadjanov:2016fbd,Briceno:2016kkp,Briceno:2015csa,Briceno:2014uqa}.}~\cite{Hoja:2010fm,Bernard:2012bi,Baroni:2018iau,Briceno:2019nns,Briceno:2020xxs}, whereas the contact term
can be dealt with in the
standard fashion. On the other hand, on the lattice, the form factor does not break
down into separate contributions. Two solutions to this problem are known in the literature.
One of the solution amounts to introducing a finite volume function which isolates the problematic triangle singularity. This is subtracted from the lattice data, leaving an amplitude that captures the short-distance physics. The infinite volume limit is then performed, and the triangle diagram contribution is added back to the final result. See Refs.~\cite{Hoja:2010fm,Bernard:2012bi,Baroni:2018iau,Briceno:2019nns,Briceno:2020vgp,Briceno:2020xxs,Moscoso:2026wmz} for more details. The second solution is based on the use of the Feynman-Hellman theorem, and was introduced in Ref.~\cite{Lozano:2022kfz}. In this method, the system is rather placed in a background field, and the energy shift is used to determine the relevant quantities.

The outline of the paper is as follows. As already mentioned, in NREFT the physically most interesting contribution arises from the contact term. Consequently, the first part of the paper focuses on this contribution, and how it is estimated in ChPT. In Sect.~\ref{sec:notions} the kinematics is defined, and a rigorous field-theoretical definition of the resonance form factors is given. The definition of polarization vectors is relegated to Appendix~\ref{app:polarization}.
In Sect.~\ref{sec:infinite} we briefly consider the fundamental concepts of the manifestly Lorentz-invariant NREFT in the context of the present problem, and present the $\rho$-meson form factors in the infinite volume. The modified L\"uscher equation, which determines the finite-volume energy levels in the presence of the background field, is derived in
Sect.~\ref{sec:levels}. Here, we also present an outline of future lattice calculations.
The numerical calculation of the form factors is presented in Sect.~\ref{sec:numerics}, and the matching of the couplings to the contact term with the low-energy constants of ChPT is deferred to Appendix~\ref{app:ChPT}.
In Sect.~\ref{sec:concl} we conclude.

\section{The form factor of a resonance in a field theory}
\label{sec:notions}

\subsection{Form factor}

To render the presentation of the material self-contained,here we collect the
basic definitions and define the relevant kinematic variables. We start with a rigorous
definition of the form-factor of a resonance,\footnote{We call this the form factor for brevity.
However, it is clear that in the case of the $\rho$-meson, we have three form factors.} which is a straightforward generalization of
the definition used in case of stable particles~\cite{Mandelstam:1955sd}. Here, we
closely follow the formulation used in Ref.~\cite{Bernard:2012bi}, adapting it to a
particular system considered (see also Refs.~\cite{Albaladejo:2012te,Djukanovic:2013mka}, where a similar method has been used in the context of ChPT calculations). Namely, let $\Psi_\mu(x)$ be a field with the quantum
numbers of the $\rho^+$-meson, obeying the constraint $\partial^\mu\Psi_\mu(x)=0$.
The two-point function
\eq\label{eq:Dmunu}
D_{\mu\nu}(P)=i\int d^4x\,e^{iPx}\langle 0|T\Psi_\mu(x)\Psi_\nu^\dagger(0)|0\rangle
\en
develops a simple pole at $P^2=s_R$ on the second Riemann sheet of the complex variable
$P^2$. In the vicinity of this pole,
\eq
D_{\mu\nu}(P)\to\frac{Z}{s_R-P^2}\,
\sum_s\varepsilon_\mu(P,s)\tilde\varepsilon_\nu(P,s)
=\frac{Z}{s_R-P^2}\,\left(-g_{\mu\nu}+\frac{P_\mu P_\nu}{P^2}\right)\, .
\en
Here, $\varepsilon_\mu(P,s)$ denotes the polarization vector for the $\rho$-meson
and
$Z$ is the wave function renormalization constant. Despite the fact that the $\rho$-meson
is not a stable particle, the polarization vector is mathematically a well-defined quantity
and obeys the relation $P^\mu\varepsilon_\mu(P,s)=0$ for complex $P^2$.
For more details and, in particular, for the definition of the ``conjugated'' vector
$\tilde\varepsilon_\mu(P,s)$ in case of an unstable particle,
see appendix~\ref{app:polarization}.

Let us now consider the three-point function
\eq
G_{\mu\nu}^\alpha(P,Q)=
\int d^4x\,d^4y\,e^{iPx-iQy}\langle 0|T\Psi_\mu(x)J^\alpha(0)\Psi_\nu^\dagger(y)|0\rangle\, ,
\en
where $J^\alpha(0)$ stands for the electromagnetic current. 
This three-point function develops a double pole at $P^2,Q^2=s_R$ in
the complex plane. In the vicinity of this double pole, the three-point function takes
the form
\eq
G_{\mu\nu}^\alpha(P,Q)=-\sum_{s,s'}
\frac{Z^{1/2}\varepsilon_\mu(P,s)}{s_R-P^2}\,
\langle P,s|J^\alpha(0)|Q,s'\rangle
\frac{Z^{1/2}\tilde\varepsilon_\nu(Q,s')}{s_R-Q^2} + \cdots\, ,
\en
where $k^\mu=(P-Q)^\mu$ and the ellipses denote less singular terms.
The above expression {\em defines} the
form factor of the $\rho$-meson $\langle P,s|J^\alpha(0)|Q,s'\rangle$
through the residue of the three-point function (here $J^\alpha(0)$ denotes the electromagnetic current). Note that this is a convenient notation only -- the one-particle state,
containing the $\rho$-meson does not appear in the Fock space.

Separating out the initial- and the final-state interactions, one may further define the amputated
three-point function
\eq\label{eq:hatG}
G_{\mu\nu}^\alpha(P,Q)=D_{\mu\sigma}(P)\hat G^{\alpha,\sigma\rho}(P,Q)
D_{\rho\nu}(Q)\, .
\en
Extracting the residue of the double pole at $P^2,Q^2\to s_R$, one gets the final expression for the
resonance form factor in the infinite volume
\eq\label{eq:limes}
\langle P,s|J^\alpha(0)|Q,s'\rangle=-Z\lim_{P^2,Q^2\to s_R}
\tilde\varepsilon_\mu(P,s)\hat G^{\alpha,\mu\nu}(P,Q)\varepsilon_\nu(Q,s')\, .
\en
This expression enables one to systematically calculate the form factor in perturbation theory,
both in the relativistic and non-relativistic settings. The final result does not depend on the
choice of the interpolating field $\Psi_\mu(x)$. Note also that this is not the case
if the form-factor is defined differently, e.g., with $P^2,Q^2$ constrained to the real axis.

Owing to the electromagnetic current conservation, the form factor obeys the relation
\eq
k_\alpha\langle P,s|J^\alpha(0)|Q,s'\rangle=0\, .
\en
It can be further shown that $\langle P,s|J^\alpha(0)|Q,s'\rangle$ can be decomposed into three linearly independent invariant form factors, which are traditionally chosen
as follows (see, e.g., \cite{Hedditch:2007ex})
\eq\label{eq:scalar_ffs}
\langle P,s|J^\alpha(0)|Q,s'\rangle
&=&\tilde\varepsilon_\mu(P,s)\Gamma^{\alpha,\mu\nu}(P,Q)
\varepsilon_\nu(Q,s')\, ,
\nonumber\\[2mm]
\Gamma^{\alpha,\mu\nu}(P,Q)&=&-G_1(k^2)g^{\mu\nu}(P+Q)^\alpha
-G_2(k^2)(k^\mu g^{\nu\alpha}-k^\nu g^{\mu\alpha})
\nonumber\\[2mm]
&&+\,\frac{1}{2s_R}\,G_3(k^2)
k^\mu k^\nu (P+Q)^\alpha\, .
\en
where, owing to the Ward identity, $G_1(k^2)$ is normalized to unity at $k^2=0$.

\subsection{Kinematics}

At the end of this section we define the kinematical variables which will be used in the following. Namely, we define the unit vectors $w^\mu$ and $u^\mu$ according to
\eq
w^\mu=\frac{P^\mu}{\sqrt{P^2}}\, ,\quad\quad
u^\mu=\frac{Q^\mu}{\sqrt{Q^2}}\, ,\quad\quad w^2=u^2=1\,.
\en
Furthermore, in the Lorentz-invariant NREFT, one needs to define the quantization
axis determined by a unit timelike vector $v^\mu$. At the end,
this vector can be fixed arbitrarily in terms of external momenta. As shown in
Ref.~\cite{Lozano:2022kfz}, it is convenient to work in the Breit frame, which for stable
particles is defined by the condition $\bm{P}=-\bm{Q}$. Here we want to generalize
this condition for a resonance form factor (corresponding to the two-particle
scattering in the presence of an background field), where $P^2$ and $Q^2$ can be treated
as independent variables. This can be achieved by defining
\eq
v^\mu=\frac{1}{2x}\,(w+u)^\mu\, ,\quad\quad x=(v\cdot w)=(v\cdot u)\, ,\quad\quad
(w\cdot u)=z=2x^2-1\, .
\en
The quantity $z$ can be expressed through external momenta
\eq
z=\frac{P^2+Q^2-k^2}{2\sqrt{P^2}\sqrt{Q^2}}\, .
\en
On the mass shell, one has to take $P^2=Q^2=s_R$. Furthermore, it is convenient to define
\eq
n^\mu=\frac{1}{2\sqrt{x^2-1}}\,(w^\mu-u^\mu)\, ,\quad\quad
n^2=-1\, .
\en
And the matrix of the Lorentz boost which renders the vector $u^\mu$ parallel to $v^\mu$
is given by
\eq
\Lambda(v,u)^{\mu\nu}&=&g^{\mu\nu}
-\frac{u^\mu u^\nu}{1+x}
-\frac{v^\mu v^\nu}{1+x}
+x\frac{u^\mu v^\nu+v^\mu u^\nu}{1+x}
-(u^\mu v^\nu-v^\mu u^\nu)\, ,
\nonumber\\[2mm]
\Lambda(v,u)^{\mu\nu}u_\nu&=&v^\mu\, .
\en
Similarly, the vector $w^\mu$ can be boosted to the rest frame defined by the vector
$v^\mu$ by using the matrix $\Lambda(v,w)^{\mu\nu}$, which is obtained from
the above expression by substituting $u\to w$.

\section{The electromagnetic form factor in the infinite volume}
\label{sec:infinite}

\subsection{The non-relativistic Lagrangian and matching in the strong sector} 

As usual, we use the NREFT framework which enables one to relate the infinite-
and finite-volume observables in the most convenient and economic way. We start
from the Lagrangian that defines elastic $\pi^+\pi^0$ scattering.
At this stage, it is convenient to define the composite fields 
\eq\label{eq:Psi}
\Psi^{(\ell)}_{\mu_1\ldots\mu_\ell}(x)=\left(\phi_+(x)\mathscr{Y}^{(\ell)}_{\mu_1\cdots\mu_\ell}(\stackrel{\leftrightarrow}{W})
 f_\ell(-\stackrel{\leftrightarrow}{W}^2)\phi_0(x)\right)\, .
\en
Here, the definition of the differential operator $\stackrel{\leftrightarrow}{W}^{\mu}$ is given by
\eq\label{eq:comp}
\left(\phi_+(x)\stackrel{\leftrightarrow}{W}^\mu\phi_0(x)\right)
=\frac{1}{2}\,\phi_+(x)\left(W^\mu\phi_0(x)\right)
-\frac{1}{2}\,\left(W^\mu\phi_+(x)\right)\phi_0(x)\, ,
\nonumber\\[2mm]
W^\mu=v^\mu W+i\partial_\perp^\mu\, ,\quad\quad
W=\sqrt{M^2+\partial_\perp^2}\, ,\quad\quad
\partial_\perp^\mu=\partial^\mu-v^\mu(v\cdot\partial)\, ,
\en
and $M$ denotes the pion mass.
The action of the operator $W^\mu$ on a product of
two field operators in the Lagrangian is defined by the chain
law
$W^\mu(\phi_1\phi_2)=(W^\mu \phi_1)\phi_2+\phi_1(W^\mu \phi_2)$. Furthermore,
the function $f_\ell(x)=f_\ell^{(0)}+xf_\ell^{(1)}+x^2f_\ell^{(2)}+\ldots$ will
be related to the $\pi\pi$ phase shift in the partial wave with the angular momentum
$\ell$
(see later).

The quantities
$\mathscr{Y}^{(\ell)}_{\mu_1\cdots\mu_\ell}(\stackrel{\leftrightarrow}{W})$
that enter Eq.~(\ref{eq:comp}) are traceless polynomials given by
\eq
\mathscr{Y}^{(0)}(\stackrel{\leftrightarrow}{W})&=&1\, ,
\nonumber\\[2mm]
\mathscr{Y}^{(1)}_{\mu}(\stackrel{\leftrightarrow}{W})&=&\stackrel{\leftrightarrow}{W}_\mu\, ,
\nonumber\\[2mm]
\mathscr{Y}^{(2)}_{\mu\nu}(\stackrel{\leftrightarrow}{W})
&=&\frac{3}{2}\,\stackrel{\leftrightarrow}{W}_{\mu}\stackrel{\leftrightarrow}{W}_{\nu}-\frac{1}{2}\,\left(g_{\mu\nu}-\frac{W_{\mu}W_{\nu}}{W^2}\right)
\stackrel{\leftrightarrow}{W}^2\, ,
\en
and so on.

In order to write down the expression for an arbitrary $\ell$, it is convenient to
first boost the momenta to the center-of-mass (CM) frame, where the
relative momentum has no zeroth component.
Then, $\mathscr{Y}^{(\ell)}_{\mu_1\cdots\mu_\ell}(\bm{k})$ vanishes, if at least one of the
indices takes the zero value $\mu_k=0$. Furthermore, for the spatial components,
\eq
\mathscr{Y}^{(\ell)}_{i_1\cdots i_\ell}(\bm{k})
=N_\ell\sum_m\sqrt{\frac{4\pi}{2\ell+1}}\,c^{\ell m}_{i_1\cdots i_\ell}\left(\mathscr{Y}_{\ell m}(\bm{k})
\right)^*\, ,\quad\quad \mathscr{Y}_{\ell m}(\bm{k})=|\bm{k}|^\ell Y_{\ell m}(\hat k)\, ,
\en
where $Y_{\ell m}(\hat k)$ stands for the conventional spherical harmonic.
This expression can be now
boosted back to an arbitrary reference frame.
Explicit expressions for the coefficients $N_\ell$ and $c^{\ell m}_{i_1\cdots i_\ell}$
are given, e.g., in Ref.~\cite{Muller:2022oyw}. Here, it suffices to know that
\eq
\sum_mc^{\ell m}_{i_1\cdots i_\ell}\left(c^{\ell m}_{i_1'\cdots i_\ell'}\right)^*
&=&\frac{1}{\ell!}\, \left(\delta_{i_1i_1'}\cdots\delta_{i_\ell i_\ell'}+\mbox{permutations}\right)\,,
\nonumber\\[2mm]
\sum_{i_1\cdots i_\ell}c^{\ell m}_{i_1\cdots i_\ell}\left(c^{\ell m'}_{i_1\cdots i_\ell}\right)^*&=&\delta^{mm'}\, .
\en
The NREFT Lagrangian is conveniently written down as
\eq
\mathscr{L}=\sum_{a=+,0}\phi^\dagger_a\,2W(i(v\cdot\partial)-W)\phi_a
+\sum_\ell\sigma_\ell\Psi^{(\ell)\dagger}_{\mu_1\cdots\mu_\ell}\Psi^{(\ell)\mu_1\cdots\mu_\ell}\, .
\en
Here, $\sigma_\ell=\pm 1$, 
according to the sign of the scattering length in the partial wave with the angular momentum $\ell$, see later.

In order to get a scattering amplitude for the process
$\pi^+(q_1)+\pi^0(q_2)\to \pi^+(p_1)+\pi^0(p_2)$, one has to sum
up an infinite chain of bubble diagrams. The on-shell amplitude in a partial wave
with angular momentum $\ell$ is given by
\eq
T_\ell(P,p,q)&=&\sigma_\ell(\mathscr{Y}^{(\ell)}(p)\cdot \mathscr{Y}^{(\ell)}(q))
f_\ell(-p^2)f_\ell(-q^2)
\nonumber\\[2mm]
&+&\int\frac{d^Dl_1}{(2\pi)^Di}\,
\frac{(\mathscr{Y}^{(\ell)}(p)\cdot \mathscr{Y}^{(\ell)}(l))
 f_\ell(-p^2)f_\ell^2(-l^2)f_\ell(-q^2)
 (\mathscr{Y}^{(\ell)}(l)\cdot \mathscr{Y}^{(\ell)}(q))}
{2W(l_1)(W(l_1)-(v\cdot l_1)-i\varepsilon)}
\nonumber\\[2mm]
&\times&\frac{1}{2W(P-l_1)(W(P-l_1)-(v\cdot (P-l_1))-i\varepsilon)}
+\cdots\, ,
\en
where
\eq
W(l_1)&=&\sqrt{M^2-l_1^2+(v\cdot l_1)^2}\, ,
\nonumber\\[2mm]
W(P-l_1)&=&\sqrt{M^2-(P-l_1)^2+(v\cdot (P-l_1))^2}\, ,
\nonumber\\[2mm]
\tilde{l}_1^\mu&=&l_1^\mu-v^\mu((v\cdot l_1)-W(l_1))\, ,
\nonumber\\[2mm]
\widetilde{(P-l_1)}^\mu&=&(P-l_1)^\mu-v^\mu((v\cdot (P-l_1))-W(P-l_1))\, ,
\nonumber\\[2mm]
l^\mu&=&\frac{1}{2}\,(\tilde{l}_1-\widetilde{(P-l_1)})^\mu\, ,
\en
and
\eq
(\mathscr{Y}^{(\ell)}(p)\cdot \mathscr{Y}^{(\ell)}(q))
=\mathscr{Y}_{\mu_1\cdots\mu_\ell}^{(\ell)}(p) \mathscr{Y}_{\mu_1\cdots\mu_\ell}^{(\ell)}(q)\, .
\en
As shown, e.g., in Ref.~\cite{Muller:2021uur}, the denominator in the above loop can be
replaced by the relativistic expression, since the difference between the two expressions
is a low-energy polynomial that vanishes in dimensional regularization.
Furthermore, in the numerator, one could replace $l^\mu$ by
$P^\mu/2-l_1^\mu$ for the same reason.
Thus, the dependence on the vector $v^\mu$ disappears and one can boost the scattering amplitude
to the center-of-mass (CM) frame. In the CM frame, only the spherical harmonics $\mathscr{Y}^{(\ell)}(\bm{l})$ depend on the direction of the three-momentum $\bm{l}$, and
\eq
&&\int d^d\bm{l}
(\mathscr{Y}^{(\ell)}(\bm{p})\cdot \mathscr{Y}^{(\ell)}(\bm{l}))
(\mathscr{Y}^{(\ell)}(\bm{l})\cdot \mathscr{Y}^{(\ell)}( \bm{q}))\Phi(|\bm{l}|)
\nonumber\\[2mm]
&=&
c_d^{(\ell)} (\mathscr{Y}^{(\ell)}(\bm{p})\cdot \mathscr{Y}^{(\ell)}(\bm{q}))
\int d^d\bm{l}|\bm{l}|^{2\ell}\Phi(|\bm{l}|)\, ,
\en
where $d=D-1$ and $\Phi(|\bm{l}|)$ is an arbitrary function depending only on the magnitude of the three-momentum $\bm{l}$, and the $c_d^ {(\ell)}$ are numerical coefficients. We can determine these coefficients for $d=3$, since this merely amounts to fixing
the renormalization prescription. In the following, we omit the subscript $d$
in this coefficient, and get $c^{(\ell)}={N_\ell^2}/({2\ell+1})$. In particular, $c^{(0)}=1$, $c^{(1)}=1/3$, and so on.
 
Evaluating now the amplitude on the mass shell with $p^2=q^2=-q_0^2=-\left({P^2}/{4}-M^2\right)$, we get
\eq
T_\ell(P,p,q)&=&\sigma_\ell(\mathscr{Y}^{(\ell)}(p)\cdot \mathscr{Y}^{(\ell)}(q))
f_\ell^2(q_0^2)\biggl\{
1+\sigma_\ell c^{(\ell)}q_0^{2\ell} f_\ell^2(q_0^2)I(q_0^2)+\cdots\biggr\}
\nonumber\\[2mm]
&=&\frac{\sigma_\ell(\mathscr{Y}^{(\ell)}(p)\cdot \mathscr{Y}^{(\ell)}(q))f_\ell^2(q_0^2)}
{1-\sigma_\ell c^{(\ell)}q_0^{2\ell} f_\ell^2(q_0^2)I(q_0^2)}
 \, ,
\en
where
\eq
I(q_0^2)&=&\int\frac{d^Dk}{(2\pi)^Di}\,
\frac{1}{(M^2-k^2)(M^2-(P-k)^2)}
\nonumber\\[2mm]
&=&\frac{iq_0}{8\pi\sqrt{P^2}}
+\mbox{low-energy polynomial}\, .
\en
We drop this polynomial as a part of the renormalization prescription used in this paper.
Then, using unitarity, it is straightforward to relate the function $f_\ell(q_0^2)$ to
the $\pi\pi$ scattering phase
\eq\label{eq:cot}
q_0\cot\delta_\ell(q_0)=
\biggl[\frac{\sigma_\ell c^{(\ell)}q_0^{2\ell} f_\ell^2(q_0^2)}{8\pi\sqrt{P^2}}\biggr]^{-1}\, .
\en
In other words, there exists a direct relation of the expansion coefficients $f_\ell^{(0)},\ldots$
and the effective-range expansion parameters in the partial wave with the angular momentum $\ell$.

\subsection{Electromagnetic interactions: gauging the one-particle Lagrangian}

Gauging of the Lagrangians containing (stable) vector particles have been considered in the literature
earlier, see, e.g.~\cite{Davoudi:2015zda} and the references therein. The difference between these papers and our present work is twofold. First, here we consider the inclusion
of the electromagnetic interactions in the covariant version of NREFT that contains derivatives in a non-linear way. Second, the $\rho$-meson in our formalism is unstable and
we work with the pion fields only.
The electromagnetic interactions will be treated to the lowest order and, hence, no kinetic terms
for the photon will be added. Gauging the pion Lagrangian containing the square root of a differential operator is not a trivial task, because the photon field depends on the space-time coordinates. Our goal can be achieved as follows. First, note that the operator $W^2$ after
gauging $\partial_\mu\to\partial_\mu-ieA_\mu$ takes the form
\eq
W^2&=&M^2+\partial^2-(v\cdot\partial)^2\to W^2+B+O(e^2)\, ,
\nonumber\\[2mm]
B&=&-ie(\partial\cdot A+A\cdot\partial)
+ie((v\cdot\partial)(v\cdot A)+(v\cdot A)(v\cdot\partial))\, .
\en
Then,
\eq
W=\sqrt{W^2}\to W+D\, ,\quad\quad
D=\int_0^\infty dt\,e^{-tW}Be^{-tW}\, .
\en
This expression simplifies if one evaluates the matrix element of both sides
in momentum space. The integral can be calculated, resulting in
\eq
\langle p|D|q\rangle=\frac{\langle p|B|q\rangle}{W(p)+W(q)}\, .
\en
The kinetic term of the pions after gauging gets an additional contribution
\eq
\sum_{a=+,0}\phi^\dagger_a\,2W(i(v\cdot\partial)-W)\phi_a\to
\sum_{a=+,0}\phi^\dagger_a\,2W(i(v\cdot\partial)-W)\phi_a+\delta\mathscr{L}+O(e^2)
\, ,
\en
with
\eq
\delta\mathscr{L}=e\phi_+^\dagger\left[(v\cdot A)W+W(v\cdot A)\right]\phi_+
+i\phi_+^\dagger\left[(v\cdot \partial)D+D(v\cdot \partial)\right]\phi_+
-2\phi_+^\dagger B\phi_+\, .
\en
The pion-photon vertex
$N^\mu(p,q)= \langle p|J^\mu(0)|q\rangle$
in momentum space can be read off from the above
Lagrangian
\eq\label{eq:Nmu}
N^\mu(p,q)&=&e(p+q)^\mu
\biggl(2-\frac{(v\cdot p)+(v\cdot q)}{W(p)+W(q)}\biggr)
\nonumber\\[2mm]
&+&ev^\mu\biggl((W(p)+W(q))+\frac{((v\cdot p)+(v\cdot q))^2}{W(p)+W(q)}
-2((v\cdot p)+(v\cdot q))\biggr)\, .
\en
On the mass shell, where $(v\cdot p)=W(p)$ and $(v\cdot q)=W(q)$, this expression 
reduces to the standard form of the vertex $\langle p|J^\mu(0)|q\rangle=e(p+q)^\mu$.
Furthermore, it is straightforward to check that the above vertex obeys the Ward
identity
\eq\label{eq:Ward}
(p-q)_\mu N^\mu(p,q)&=&eS^{-1}(q)-eS^{-1}(p)\, ,
\nonumber\\[2mm]
S^{-1}(p)&=&2W(p)(W(p)-(v\cdot p))\, .
\en
At higher orders, one has to add terms that reproduce the expansion of the charged
pion form factor $F_\pi(k^2)$ in powers of $k^2=(p-q)^2$. It suffices to multiply the whole vertex
with $F_\pi(k^2)$, since all other terms yield vanishing contributions on the mass shell.

One can further argue that there is no need to consider gauging the derivative terms in the
four-pion interaction Lagrangian. This procedure will yield local terms in the Lagrangian containing four pion and one photon field. Since we are going to add such terms anyway, gauging the four-pion Lagrangian is superfluous.

\subsection{Electromagnetic interactions: five-point Green functions}

As already mentioned, at order $e$, the gauged strong Lagrangian should be amended
by local counterterms containing four pion fields and the photon. In writing
down these counterterms, it is convenient to
use the fields $\Psi^{(\ell)}_{\mu_1\ldots\mu_\ell}(x)$, introduced in Eq.~(\ref{eq:Psi}).
Since the Lagrangian must be gauge-invariant, it should contain the
quantity $F_{\alpha\beta}=\partial_\alpha A_\beta-\partial_\beta A_\alpha$
instead of the electromagnetic field $A_\alpha$. For convenience, we further define
$B_\alpha=\partial^\beta F_{\alpha\beta}$, which is also a gauge-invariant quantity.

We start from the Lagrangian in which both the incoming and the outgoing pion pairs have
angular momentum $\ell'=\ell=1$. In the infinite volume, this is the only part of the
Lagrangian that contributes to the $\rho$-meson form factor. At the lowest order, the counterterm Lagrangian contains four terms
\eq\label{eq:Lct}
\mathscr{L}_{\sf c.t.}=\sum_{i=1}^4g_ie\mathscr{O}_i\, ,
\en
where
\eq\label{eq:g123}
\mathscr{O}_1&=&\Psi^{(1)\dagger}_\mu\{B^\alpha,W_\alpha\}g^{\mu\nu}\Psi^{(1)}_\nu\, ,
\nonumber\\[2mm] 
\mathscr{O}_2&=&i\Psi^{(1)\dagger}_\mu F^{\mu\nu}\Psi^{(1)}_\nu\, ,
\nonumber\\[2mm] 
\mathscr{O}_3&=&\frac{1}{2}\,\Psi^{(1)\dagger}_\mu\{(\partial^\mu F^{\nu\alpha}+\partial^\nu F^{\mu\alpha}),W_\alpha\}\Psi^{(1)}_\nu\, ,
\nonumber\\[2mm] 
\mathscr{O}_4&=&\Psi^{(1)\dagger}_\mu[W^\mu,[W^\nu,\{B^\alpha,W_\alpha\}]]\Psi^{(1)}_\nu\, .
\en
In the following, it will be convenient to work in momentum space. Let us consider
the scattering process $\pi^+(q_1)+\pi^0(q_2)+A^\alpha(k)\to \pi^+(p_1)+\pi^0(p_2)$.
The CM and relative momenta are defined as
\eq\label{eq:CM-relative}
P&=&p_1+p_2\, ,\quad
Q=q_1+q_2\, ,
\nonumber\\[2mm]
p&=&\frac{1}{2}\, (p_1-p_2)\, ,\quad
q=\frac{1}{2}\,(q_1-q_2)\, ,
\nonumber\\[2mm]
k&=&P-Q\, .
\en
The tree-level contribution of the above Lagrangian to the matrix element of the electromagnetic current between the two-pion states can be evaluated straightforwardly:
\eq\label{eq:matrixelement}
&&\langle\pi^+(p_1)\pi^0(p_2)|J_\alpha(0)|\pi^+(q_1)\pi^0(q_2)\rangle_{\sf c.t.}
\nonumber\\[2mm]
&=&
g_1ef(-p^2)f(-q^2)(p\cdot q)(k^2g_{\alpha\beta}-k_\alpha k_\beta)(P+Q)^\beta
\nonumber\\[2mm]
&+&g_2ef(-p^2)f(-q^2)p^\mu q^\nu(k_\mu g_{\nu\alpha}-k_\nu g_{\mu\alpha})
\nonumber\\[2mm]
&+&g_3ef(-p^2)f(-q^2)\biggl((p\cdot k)(q\cdot k)(P+Q)^\alpha-
\frac{1}{2}\,((p\cdot k)q^\alpha+(q\cdot k)p^\alpha)(P^2-Q^2)
\biggr)
\nonumber\\[2mm]
&+&g_4ef(-p^2)f(-q^2)(p\cdot k)(q\cdot k) (k^2g_{\alpha\beta}-k_\alpha k_\beta)(P+Q)^\beta\, .
\en
In general, there exist six independent structures that are linear in momenta $p$ and $q$:
\eq
S^{(1)}_\alpha&=&-(p\cdot q)(P+Q)_\alpha\, ,
\nonumber\\[2mm]
S^{(2)}_\alpha&=&-(k\cdot p)q_\alpha+(k\cdot q)p_\alpha\, ,
\nonumber\\[2mm]
S^{(3)}_\alpha&=&(k\cdot p)(k\cdot q)(P+Q)_\alpha\, ,
\nonumber\\[2mm]
S^{(4)}_\alpha&=&-(p\cdot q)k_\alpha(P^2-Q^2)\, ,
\nonumber\\[2mm]
S^{(5)}_\alpha&=&((k\cdot p)q_\alpha+(k\cdot q)p_\alpha)(P^2-Q^2)\, ,
\nonumber\\[2mm]
S^{(6)}_\alpha&=&(k\cdot p)(k\cdot q)k_\alpha(P^2-Q^2)\, ,
\en
Note that under the crossing transformation $p\to -q$ and $P\to -Q$, all the above structures
transform as $S^{(i)}_\alpha\to -S^{(i)}_\alpha\, ,~i=1,\ldots,6$, as expected.\footnote{This can be shown by using $C$-invariance
(recall that the photon field $A^\alpha$ is odd under charge conjugation).}
The four independent structures in Eq.~(\ref{eq:matrixelement}) are certain linear combinations
of $S_\alpha^{(1)},\cdots,S_\alpha^{(6)}$ which vanish automatically when multiplied
by $k^\alpha$. Gauge invariance imposes no further constraints
on the couplings $g_1,g_2,g_3,g_4$ exactly for this reason.

Carrying out the matching in this Lagrangian is a subtle enterprise. Consider first the matching in the general kinematics $P^2\neq Q^2$. The number of independent couplings
is equal to four. However, the constraints imposed by gauge invariance hold for the
whole matrix element of the electromagnetic current and not separately for the loop
contribution or the local counterterm contribution. Hence, in principle, if the loop alone
does not obey the Ward identity, the gauge-non-invariant local vertices will be
needed along with the four operators in Eq.~(\ref{eq:Lct}) in order to guarantee gauge invariance. The coefficients of these additional operators are however fixed.\footnote{A trivial example is provided by the relativistic scalar field theory with derivative interactions,
e.g. $\mathscr{L}_{\sf int}=g|\varphi|^2|\partial\varphi|^2$. Gauging this Lagrangian
by using $\partial_\mu\varphi\to\partial_\mu\varphi-ieA_\mu\varphi$, one arrives at the five-point local vertices that contain $A_\mu$ and are hence not explicitly gauge-invariant.}
However, as it will be shown below, the loop contribution in NREFT is gauge invariant by
itself. Hence, the additional contributions are not needed in this case.
In particular, it can be seen that there is no need to consider gauging the four-pion Lagrangian
with derivative couplings -- it can produce only the gauge-invariant contributions that
are already taken into account.

Furthermore, since we are interested in the study of the $\rho$-meson only,
we restrict our matching by the condition $P^2=Q^2$. This reduces the number of the
independent operators by one, since $\mathscr{O}_3$ and $\mathscr{O}_4$ yield
the same amplitude in this kinematics up to an overall factor $k^2$. We use this fact
and set $g_4=0$ from the beginning.

Our next remark concerns partial-wave mixing with either $\ell\neq 1$ or $\ell'\neq 1$.
Owing to rotational invariance, there is no partial-wave mixing in the infinite volume.
However, the situation changes in a finite volume, where non-diagonal transitions
between different $\ell,\ell'$ are possible. A completely general treatment of this problem
could be quite cumbersome, albeit relatively straightforward. One circumstance, however,
renders the problem much simpler: the $\rho$-meson appears in the channel with total
isospin $I=1$, where S- and D-waves do not contribute.\footnote{For simplicity, we assume the equal-mass case here and neglect the small isospin breaking due to the
charged and neutral pion mass difference.} Therefore, the mixing occurs first
in the F-wave which is very small and can be safely neglected, along with higher partial
waves.

Finally, it should be pointed out that Eqs.~(\ref{eq:Lct}), (\ref{eq:g123})
correspond to the lowest-order Lagrangian only. Higher-order terms come in a form of
Taylor expansion in the variables $p^2,q^2,k^2$. In this paper, we limit ourselves to
the lowest order only that is expected to be a good approximation in the view of the
fact that $\rho$-meson is a rather narrow resonance and the expansion in $p^2,q^2$
in the vicinity of the real part of the resonance energy is likely to converge very rapidly.
Moreover, since the lattice fit is performed separately for each value of $k^2$, the issue
of convergence in this variable never arises -- the couplings $g_1,g_2,g_3$ can be merely assumed to be the functions of $k^2$.
An order-of magnitude estimate of these couplings can be carried out in ChPT,
see appendix~\ref{app:ChPT} for more details. 

\subsection{Calculation of the $Z$-factor}

The interpolating field of the $\rho$-meson can be chosen in the following way
\eq
\Psi_\mu(x)&=&i\left(g_{\mu\nu}-\frac{\partial_\mu\partial_\nu}{\partial^2}\right)
\phi_+(x)\!\stackrel{\leftrightarrow}{\partial}^{\,\nu}\!\!f(-\stackrel{\leftrightarrow}{W}^2)\phi_0(x)\, ,
\nonumber\\[2mm]
i\phi_+(x)\!\stackrel{\leftrightarrow}{\partial}^{\,\nu}\!\!\phi_0(x)
&=& \frac{i}{2}\,\left(\phi_+(x)(\partial^\nu\phi_0(x))
-(\partial^\nu\phi_+(x))\phi_0(x)\right)\, ,
\en
where $f(x)\doteq f_1(x)$.
Note that this expression differs from the one given in Eq.~(\ref{eq:Psi}) (for $\ell=1$)
if momenta of the constituent particles are not restricted on the mass shell. Unlike
the field from Eq.~(\ref{eq:Psi}), the field $\Psi_\mu(x)$ obeys the condition
$\partial^\mu\Psi_\mu(x)=0$ by construction.

Summing up the bubbles, one arrives at the following expression for the two-point function
on the first Riemann sheet
\eq\label{eq:prop}
D_{\mu\nu}(P)&=&\left(g_{\mu\nu}-\frac{P_\mu P_\nu}{P^2}\right)D(P^2)\, ,
\nonumber\\[2mm]
D(P^2)&=&-\frac{c^{(1)}q_0^2f^2(q_0^2)I(q_0^2)}{1+c^{(1)}q_0^2f^2(q_0^2)I(q_0^2)}=\frac{iq_0}{q_0\cot\delta_1(q_0)-iq_0}\, .
\en
Here, we have used $\sigma_1=-1$, since the experimental value of the
P-wave scattering length is negative, and $q_0^2=P^2/4-M^2$. 

On the second Riemann sheet, $iq_0$ should be replaced by
$\left(M^2-{P^2}/{4}\right)^{1/2}=(-q_0^2)^{1/2}$. Expanding in the vicinity of the resonance pole $P^2=s_R$, one finally gets
\eq
Z&=&\frac{8q_R^2}{1+2(-q_R^2)^{1/2}b}\, ,
\nonumber\\[2mm]
b&=&\frac{d}{dq_0^2}\,\left[q_0\cot\delta_1(q_0)\right]\biggr|_{q_0^2=q_R^2}
\nonumber\\[2mm]
&=&\frac{1}{(-q_R^2)^{1/2}f^2(q_R^2)}\,
\left(q_R^2\frac{d}{dq_0^2}\,\left[f^2(q_0^2)\right]\biggr|_{q_0^2=q_R^2}
 +\left(1-\frac{2q_R^2}{s_R}\right)f^2(q_R^2)\right)\, .
\en
Thus, the resonance $Z$-factor can be evaluated using reliable parameterizations of
the P-wave $\pi\pi$ scattering phase shift in the resonance region.

\subsection{The $\rho$-meson form factor: the infinite volume case}

\begin{figure}[t]
\begin{center}
\includegraphics[width=14cm]{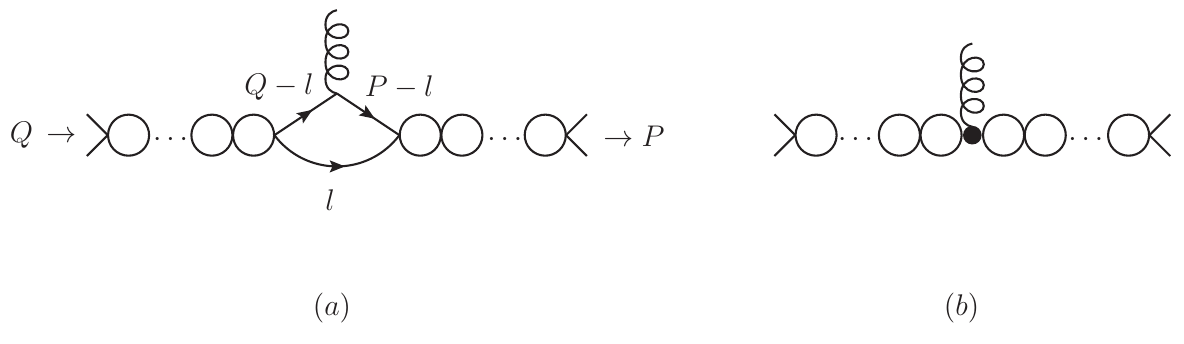}
\caption{Contributions to the rho-meson form factor in NREFT: (a) The so-called triangle
diagram, where the external photon is hooked on to the charged pion, and (b) the external photon emitted from the local five-point vertex. Both types of diagrams are dressed by an infinite number of two-pion bubbles describing the initial- and the final-state interactions.}
\label{fig:formfactor}
\end{center}
\end{figure}

In NREFT, the rho-meson form factor receives contributions from the diagrams of only
two types, which are shown in Fig.~\ref{fig:formfactor}. The diagram (a) is expressed
in terms of the known quantities only: the $\pi\pi$ phase shift in the P-wave, and the
electromagnetic form factor of a pion. In order to calculate this contribution, no
lattice input about the $\rho$-meson form factor is needed. The only unknown quantities
at lowest order are the lowest-order couplings $g_1,g_2,g_3$ that appear in the diagram (b). These are extracted
on the lattice from the finite-volume energy levels in the background field and are,
at the next stage, substituted into the infinite-volume expression. At the end, one arrives
at the $\rho$-meson form factor one is looking for.

We start by calculating the quantity $\hat G^{\alpha,\sigma\rho}(P,Q)$ which is defined
in Eq.~(\ref{eq:hatG}) in perturbation theory. An infinite sum of bubbles in the
external legs can be summed up, yielding two terms corresponding to the diagrams
of type (a) and (b) in Fig.~\ref{fig:formfactor}:
\eq
\hat G^{\alpha,\sigma\rho}(P,Q)
=\hat G_a^{\alpha,\sigma\rho}(P,Q)+\hat G_b^{\alpha,\sigma\rho}(P,Q)\, .
\en
Here,
\eq\label{eq:Gab}
\hat G_a^{\alpha,\sigma\rho}(P,Q)&=&J^{\alpha,\sigma\rho}(P,Q)\, ,
\nonumber\\[2mm]
\hat G_b^{\alpha,\sigma\rho}(P,Q)&=&
-g_1ek^2(P+Q)^\alpha g^{\sigma\rho}-g_2e(k^\sigma g^{\rho\alpha}-k^\rho g^{\sigma\alpha})
-g_3e(P+Q)^\alpha k^\sigma k^\rho\, ,\quad\quad
\en
where
\eq\label{eq:J}
J^{\alpha,\sigma\rho}(P,Q)&=&-\int \frac{d^Dl}{(2\pi)^Di}\,
\frac{\widetilde W^\sigma_P f_P N^\alpha f_Q \widetilde W^\rho_Q}
{2W(W-(v\cdot l))2W_1(W_1-(v\cdot(P-l)))}
\nonumber\\[2mm]
&\times&\frac{1}{2W_2(W_2-(v\cdot(Q-l)))}\, .
\en
Here, the following notations are used
\eq
W=W(l)\, ,\quad\quad
W_1=W(P-l)\, ,\quad\quad
W_2=W(Q-l)\, ,
\en
and
\eq
\widetilde W_P^\sigma = \frac{1}{2}\,(\widetilde{(P-l)}-\widetilde{l}\,)^\sigma\, ,\quad\quad
\widetilde W_Q^\rho = \frac{1}{2}\,(\widetilde{(Q-l)}-\widetilde{l}\,)^\rho\, ,\quad\quad
N^\alpha=N^\alpha(P-l,Q-l)\, .\quad\quad
\en
Furthermore, for any given vector
$\widetilde{p}^{\,\mu}=v^\mu W(p)+p^\mu-v^\mu(v\cdot p)=v^\mu W(p)+p_\perp^\mu$,
the vertex $N^\alpha$ is defined in Eq.~(\ref{eq:Nmu}), and
the functions $f_P,f_Q$ depend on the scalar arguments
$-(\widetilde W^\mu_P)^2$, $-(\widetilde W^\mu_Q)^2$, respectively.
Note also that we have dropped the off-shell terms that arise
owing to the difference between $\Psi_\mu$ and $\Psi^{(1)}_\mu$. These terms do not
possess a double pole and, therefore, do not contribute to the $\rho$-meson form factor.
The latter can be straightforwardly evaluated by substituting
the above expressions into Eq.~(\ref{eq:limes}).

\subsection{Ward identities}

It is easy to verify that the counterterm contribution to the form factor is transverse
by construction
\eq
k_\alpha \hat G_b^{\alpha,\sigma\rho}(P,Q)=0\, ,\quad\quad\mbox{for}\,\, P^2=Q^2\, .
\en
Using Eq.~(\ref{eq:Ward}), it can be shown that the contribution from the first term
also vanishes. Indeed,
\eq
k_\alpha J^{\alpha,\sigma\rho}(P,Q)&=&-e\int \frac{d^Dl}{(2\pi)^Di}\,
\frac{\widetilde W^\sigma_P f_P
\widetilde W^\rho_Q f_Q}
{2W(W-(v\cdot l))2W_1(W_1-(v\cdot(P-l)))}
\nonumber\\[2mm]
&+&e\int \frac{d^Dl}{(2\pi)^Di}\,
\frac{\widetilde W^\sigma_P f_P
\widetilde W^\rho_Q f_Q}
{2W(W-(v\cdot l))2W_2(W_2-(v\cdot(Q-l)))}\, .
\nonumber\\
\en
On the other hand, $k_\alpha J^{\alpha,\sigma\rho}(P,Q)$ is a second-order Lorentz-tensor,
which obeys the condition
\eq\label{eq:anti}
k_\alpha J^{\alpha,\sigma\rho}(P,Q)
=-k_\alpha J^{\alpha,\rho\sigma}(Q,P)\, .
\en
It is given as a linear combination of the invariant form factors
\eq 
k_\alpha J^{\alpha,\sigma\rho}(P,Q)
=g^{\sigma\rho}X_1+v^\sigma v^\rho X_2+n^\sigma n^\rho X_3+(v^\sigma n^\rho+v^\rho n^\sigma)X_4+(v^\sigma n^\rho-v^\rho n^\sigma)X_5\, ,
\nonumber\\
\en
with $X_i=X_i(k^2)$. Eq.~(\ref{eq:anti}) implies that $X_1=X_2=X_3=X_5=0$. Furthermore, sandwiching the remaining term with the polarization vectors and using $P^2=Q^2$,
we have
\eq
&&\tilde\varepsilon_\sigma(P,s)(v^\sigma n^\rho+v^\rho n^\sigma)\varepsilon_\rho(Q,s')
\nonumber\\[2mm]
&=&\frac{1}{4P^2x\sqrt{x^2-1}}\,
\tilde\varepsilon_\sigma(P,s)((P+Q)^\sigma (P-Q)^\rho+(P+Q)^\rho (P-Q)^\sigma)\varepsilon_\rho(Q,s')=0\, .\quad\quad
\en
Thus, the triangle loop contribution is also transverse by itself.

Owing to the gauge invariance, the form factor should be normalized to unity at zero momentum transfer. Using Lorentz invariance, we can limit ourselves to the frame
$P^\mu=Q^\mu=(P^0,\bm{0})$, $v^\mu=(1,0)$.
In this frame, 
$\frac{1}{2}\,(\widetilde{(P-l)}-\widetilde{l}\,)^\mu=(0,-\bm{l})$
and 
$\frac{1}{2}\,(\widetilde{(Q-l)}-\widetilde{l}\,)^\mu=(0,-\bm{l})$.
Hence, $J^{\alpha,00}(P,Q) =J^{\alpha,0i}(P,Q)= J^{\alpha,i0}(P,Q)=0$, and
\eq
J^{\alpha,ij}(P,Q)=-\int\frac{d^Dl}{(2\pi)^Di}\,
\frac{f^2(\bm{l}^2)\bm{l}^i \bm{l}^j N^\alpha(P-l,P-l)}
{2W(l)(W(l)-l_0)(2W(l)(W(l)-P_0+l_0))^2}\, .
\en
Furthermore,
\eq
N^0(P-l,P-l)&=&2eW(l)\, ,
\nonumber\\[2mm]
N^k(P-l,P-l)&=&-2e\frac{l^k}{W(l)}\,(2W(l)-P_0+l_0)\, .
\en
The integral over the space component vanishes, since the integrand is antisymmetric with respect to $\bm{l}\to-\bm{l}$. Carrying out contour integration over $l_0$ and averaging
over the directions of the three-momentum $\bm{l}$, one gets
\eq
J^{0,ij}(P,Q)=-\frac{e}{3}\delta^{ij}\int\frac{d^d\bm{l}}{(2\pi)^d}\,
\frac{f^2(\bm{l}^2)\bm{l}^2}
{(2W(l))^2(2W(l)-P_0)^2}\, .
\en
Expanding $f^2(\bm{l}^2)$ in the numerator, one gets on the second sheet
\eq
J^{0,ij}(P,Q)&=&-\frac{e}{3}\delta^{ij}\left(
f^2(q_R^2)I_1+\frac{d}{dq_0^2}\,\left[f^2(q_0^2)\right]\biggr|_{q_0^2=q_R^2}\,I_2
\right)
\nonumber\\[2mm]
&=&-\delta^{ij}\frac{e(-q_R^2)^{1/2}}{96\pi}\,f^2(q_R^2)(1+2(-q_R^2)^{1/2}b)\, ,
\en
where
\eq
I_1=\frac{(-q_R^2)^{1/2}}{32\pi}\,\left(3-\frac{4q_R^2}{P^2}\right)\, ,\quad\quad
I_2=\frac{(-q_R^2)^{1/2}}{16\pi}\,q_R^2\, .
\en
Putting the pieces together, one gets
\eq
\langle P,s|J^0(0)|P,s'\rangle=eZ\delta_{ss'}\,\frac{e(-q_R^2)^{1/2}}{96\pi}\,f^2(q_R^2)
(1+2(-q_R^2)^{1/2}b)=2P_0\,e\delta_{ss'}\, .
\en
As expected, the form factor is normalized to unity at zero momentum transfer.

\subsection{Invariant form factors}

Let us introduce a useful notation (cf. with Eq.~(\ref{eq:J})):
\eq
\langle X\rangle
=-\int \frac{d^Dl}{(2\pi)^Di}\,
\frac{f_P X f_Q}
{2W(W-(v\cdot l))2W_1(W_1-(v\cdot(P-l)))2W_2(W_2-(v\cdot(Q-l)))}\, .
\nonumber\\
\en
Then,
\eq\label{eq:langle-rangle}
\langle P,s|J^\alpha(0)|Q,s'\rangle=
Z\tilde\varepsilon_\mu(P,s)\left\{-\hat G_b^{\alpha,\mu\nu}(P,Q)+
\langle \widetilde W^\mu_P\widetilde W^\nu_Q N^\alpha\rangle\right\}\varepsilon_\nu(Q,s')\, ,
\en
where $\hat G_b$ is given in Eq.~(\ref{eq:Gab}), and the vectors $\widetilde W^\mu_P$,
$\widetilde W^\nu_Q$ can be rewritten in the following form 
\eq
\widetilde W^\mu_P 
=-l_\perp^\mu+\frac{1}{2}\,v^\mu\Delta_P\, ,\quad\quad
\widetilde W^\nu_Q 
=-l_\perp^\nu+\frac{1}{2}\,v^\nu\Delta_Q\, ,
\en
Note that in the above expressions for $\widetilde W^\mu_P$, $\widetilde W^\nu_Q$ we have
dropped the terms proportional to $P^\mu$ and $Q^\nu$, respectively, since these are
convolved with the corresponding polarization vectors and do not contribute.
Here, $l_\perp^\mu=l^\mu-v^\mu(v\cdot l)$, and
\eq
\Delta_P&=&W_1-W-(v\cdot(P-l))-(v\cdot l)\, ,
\nonumber\\[2mm]
\Delta_Q&=&W_2-W-(v\cdot(Q-l))-(v\cdot l)\, ,
\nonumber\\[2mm]
\delta&=&-2 \biggl(2-\frac{(v\cdot (P-l))+(v\cdot (Q-l))}{W_1+W_2}\biggr)\, ,
\nonumber\\[2mm]
\Delta&=&2(x\sqrt{s_R}-(v\cdot l)) \biggl(2-\frac{(v\cdot (P-l))+(v\cdot (Q-l))}{W_1+W_2}\biggr)
\nonumber\\[2mm]
&+&
\frac{\bigl(W_1+W_2-(v\cdot (P-l))-(v\cdot (Q-l))\bigr)^2}{W_1+W_2}\, .
\en
It is convenient to carry out the calculations in the Breit frame, where one may use the
symmetries of the integrand with respect to $P^\mu\leftrightarrow Q^\mu$ and
$l_\perp^\mu\to -l_\perp^\mu$. Owing to these symmetries, some of the integrals vanish.
In general, the quantity $\langle \widetilde W^\mu_P\widetilde W^\nu_Q N^\alpha\rangle$
which enters the expression of the form factors (see Eq.~(\ref{eq:langle-rangle})),
is a Lorentz tensor of rank 3, which can be built out of the available
vectors $v^\mu$, $n^\mu$ and
the metric tensor $g^{\mu\nu}$. The numerator of this expression contains the vectors
$l_\perp^\mu$ and $v^\mu$ only. The list of all possible structures consistent with the symmetries is thus given by
\eq
\langle l_\perp^\mu l_\perp^\nu l_\perp^\alpha\varphi\rangle&=&0\, ,
\nonumber\\[2mm]
\langle l_\perp^\mu l_\perp^\nu \varphi\rangle&=&
(g^{\mu\nu}-v^\mu v^\nu) A_1+n^\mu n^\nu B_1\, ,
\nonumber\\[2mm]
v^\mu \langle l_\perp^\nu l_\perp^\alpha\varphi^{(1)}\rangle
+v^\nu \langle l_\perp^\mu l_\perp^\alpha\varphi^{(2)}\rangle &=&
v^\mu\left((g^{\nu\alpha}-v^\nu v^\alpha)A_2^{(1)}+n^\nu n^\alpha B_2^{(1)}\right)
\nonumber\\[2mm]
&-&v^\nu\left((g^{\mu\alpha}-v^\mu v^\alpha)A_2^{(2)}+n^\mu n^\alpha B_2^{(2)}\right)\, ,
\nonumber\\[2mm]
\langle l_\perp^\alpha\varphi\rangle&=&0\, ,
\nonumber\\[2mm]
v^\mu v^\alpha \langle l_\perp^\nu \varphi^{(1)}\rangle
+v^\nu v^\alpha \langle l_\perp^\mu \varphi^{(2)}\rangle
&=&v^\mu v^\alpha n^\nu A_3^{(1)}+v^\nu v^\alpha n^\mu A_3^{(2)}\, .
\en
Here, $\varphi$ denotes any Lorentz-invariant scalar function that is symmetric
with respect to $P^\mu \leftrightarrow Q^\mu$, whereas $\varphi^{(1)}$ and $\varphi^{(2)}$ transform into each other in a result of such a replacement (the same for the invariant
functions $A_i,B_i,C_i,D_i$). We furthermore
use the fact that the polarization vectors are orthogonal to the corresponding
momenta and obtain
\eq
n^\mu \tilde\varepsilon_\mu(P,s)=-\frac{x}{\sqrt{x^2-1}}\, v^\mu \tilde\varepsilon_\mu(P,s)\, ,
\quad\quad
n^\nu \varepsilon_\nu(Q,s')=\frac{x}{\sqrt{x^2-1}}\, v^\nu \varepsilon_\nu(Q,s')\, .
\en
It is then straightforward to single out the invariant form factors, see Eq.~(\ref{eq:scalar_ffs})
\eq\label{eq:final-G}
G_1(k^2)&=&-eZg_1k^2+eZ\langle F_1\rangle\, ,
\nonumber\\[2mm]
G_2(k^2)&=&-eZg_2+eZ\langle F_2\rangle\, ,
\nonumber\\[2mm]
G_3(k^2)&=&2es_RZg_3+eZ\langle F_3\rangle\, ,
\en
where
\eq
F_1&=&-\frac{1}{4x\sqrt{s_R}}\,(l_\perp^2+(n\cdot l_\perp)^2)\Delta\, ,
\nonumber\\[2mm]
F_2&=&-\frac{1}{16x\sqrt{s_R}}\,
(l_\perp^2+(n\cdot l_\perp)^2)(\Delta_P+\Delta_Q)\delta\, ,
\nonumber\\[2mm]
F_3&=&\frac{1}{16x^3\sqrt{s_R}}\biggl\{
-\frac{1}{4}\,(l_\perp^2+(n\cdot l_\perp)^2)(2\Delta-(\Delta_P+\Delta_Q)\delta)
+\frac{x^2}{2(x^2-1)}\,(l_\perp^2+3(n\cdot l_\perp)^2)\Delta
\nonumber\\[2mm]
&+&\frac{x}{2\sqrt{x^2-1}}\,(n\cdot l_\perp)(\Delta_P-\Delta_Q)\Delta
+\frac{1}{4}\,\Delta_P\Delta_Q\Delta\biggr\}\, .
\en
The calculations are most easily done in the Breit frame.
In three dimensions, we choose $\bm{P}=(0,0,|\bm{P}|)$, $\bm{Q}=(0,0,-|\bm{P}|)$,
$P_0=Q_0=\sqrt{s_R+\bm{P}^2}$
and obtain $v^\mu=(1,0,0,0)$, $n^\mu=(0,0,0,1)$. In $D$ dimensions,
only the third spatial component of the vectors $\bm{P},\bm{Q},\bm{n}$ is nonzero.
After performing the Cauchy integral and using some straightforward algebra, the
integrals that enter the expression of the form factors are given by
\eq
&&\quad \langle F_i\rangle=\int\frac{d^d\bm{l}}{(2\pi)^d}\,\Phi_i(l_3,\bm{l}_{\sf tr}^2)
\nonumber\\[2mm]
&\times&\frac{1}
{\left(\dfrac{s_R}{P_0^2}\,l_3^2-\dfrac{s_R|\bm{P}|}{P_0^2}\,l_3-\dfrac{s_R^2}{4P_0^2}
+\bm{l}_{\sf tr}^2+M^2\right)
\left(\dfrac{s_R}{P_0^2}\,l_3^2+\dfrac{s_R|\bm{P}|}{P_0^2}\,l_3-\dfrac{s_R^2}{4P_0^2}
+\bm{l}_{\sf tr}^2+M^2\right)}\, ,\quad\quad
\en
where $i=1,2,3$. Furthermore, $\bm{l}=(\bm{l}_{\sf tr},l_3)$ and $d^d\bm{l}=d^{d-1}\bm{l}_{\sf tr}dl_3$. The functions $\Phi_i$, which appear in the numerator, are given by
\eq
\Phi_i(l_3,\bm{l}_{\sf tr}^2)&=&\bar F_i(l_3,\bm{l}_{\sf tr}^2)
f\left(\left(\frac{\bm{P}}{2}-\bm{l}\right)^2-\frac{1}{4}\,(W_1-W)^2\right)
f\left(\left(\frac{\bm{Q}}{2}-\bm{l}\right)^2-\frac{1}{4}\,(W_2-W)^2\right)
\nonumber\\[2mm]
&\times&\frac{(W+W_1+P_0)((W-W_1)^2-P_0^2)(W+W_2+P_0)((W-W_2)^2-P_0^2)}{16P_0^42W2W_12W_2}
\, ,\quad\quad
\en
where
\eq\label{eq:barF}
\bar F_1&=&\frac{1}{4x\sqrt{s_R}}\,(\bm{l}^2-(\bm{n}\cdot \bm{l})^2)\bar \Delta\, ,
\nonumber\\[2mm]
\bar F_2&=&\frac{1}{16x\sqrt{s_R}}\,(\bm{l}^2-(\bm{n}\cdot \bm{l})^2)(\bar\Delta_P+\bar\Delta_Q)\bar\delta\, ,
\nonumber\\[2mm]
\bar F_3&=&\frac{1}{16x^3\sqrt{s_R}}\biggl\{
\frac{1}{4}\,(\bm{l}^2-(\bm{n}\cdot \bm{l})^2)(2\bar\Delta-(\bar\Delta_P+\bar\Delta_Q)\bar\delta)
-\frac{x^2}{2(x^2-1)}\,(\bm{l}^2-3(\bm{n}\cdot \bm{l})^2)\bar\Delta
\nonumber\\[2mm]
&-&\frac{x}{2\sqrt{x^2-1}}\,(\bm{n}\cdot \bm{l})(\bar\Delta_P-\bar\Delta_Q)\bar\Delta
+\frac{1}{4}\,\bar\Delta_P\bar\Delta_Q\bar\Delta\biggr\}\, .
\en
In the above expressions, the following notations are used:
\eq
\bar\Delta_P&=&W_1-W-P_0\, ,
\nonumber\\[2mm]
\bar\Delta_Q&=&W_2-W-P_0\, ,
\nonumber\\[2mm]
\bar\delta&=&-4 \biggl(1-\frac{P_0-W}{W_1+W_2}\biggr)\, ,
\nonumber\\[2mm]
\bar\Delta&=&4(x\sqrt{s_R}-W) \biggl(1-\frac{P_0-W}{W_1+W_2}\biggr)
+\frac{(W_1+W_2+2W-2P_0)^2}{W_1+W_2}\, .
\en
Rescaling $l_3\to \frac{P_0}{\sqrt{s_R}}\,l_3$,
and using Feynman parameterization in the denominator and the symmetry of the numerator
with respect to $l_3\to -l_3$, we get:
\eq
\langle F_i\rangle=\frac{1}{2}\,\int_0^{1} du
\int\frac{d^d\bm{l}}{(2\pi)^d}\,
\frac{\hat\Phi_i(u;\l_3^2,\bm{l}_{\sf tr}^2)}
{(\bm{l}^2-z^2)^2}\, ,
\en
where
\eq
\hat\Phi_i(u;\l_3^2,\bm{l}_{\sf tr}^2)=
\frac{P_0}{\sqrt{s_R}}\,\left(
\Phi_i\left(\frac{P_0}{\sqrt{s_R}}\,l_3-\frac{|\bm{P}|}{2}\,(1-2u),\bm{l}_{\sf tr}^2\right)
+ \Phi_i\bigl( u\to (1-u) \bigr)\right)\, .
\en
and
\eq
z^2=\frac{s_R}{4}-M^2-\frac{\bm{P}^2s_R}{P_0^2}\,u(1-u)\, .
\en
At this stage, it is convenient to introduce polar coordinates in the $d$-dimensional space
\eq
l_3&=&|\bm{l}|\cos\theta\, ,\quad\quad
|\bm{l}_{\sf tr}|=|\bm{l}|\sin\theta\, ,\quad\quad
\nonumber\\[2mm]
d^d\bm{l}&=&dl_3d|\bm{l}_{\sf tr}||\bm{l}_{\sf tr}|^{d-2}d\Omega_{d-1}
=d\theta d|\bm{l}||\bm{l}|^{d-1}\sin\theta^{d-2}d\Omega_{d-1}\, ,
\en
where $d\Omega_{d-1}$ denotes the measure of the angular integral in the $d-1$ dimensional space orthogonal to the first axis, and the angle $\theta$ runs from $0$ to $\pi$. The
integrand does not depend on the angles contained in $d\Omega_{d-1}$ and, hence, the
integral over these angles can be evaluated, yielding the volume of the $d-1$-dimensional unit sphere $\Omega_{d-1}$.

Furthermore, the numerator can be expanded in a Taylor series in $(\bm{l}^2-z^2)$:
\eq
\hat\Phi_i(u;\bm{l}^2\cos^2\theta,\bm{l}^2\sin^2\theta)=\Phi^{(0)}_i(u;\theta)
+(\bm{l}^2-z^2)\Phi^{(1)}_i(u;\theta)+O((\bm{l}^2-z^2)^2)\, .
\en
According to the rules of dimensional regularization, the higher-order terms do not contribute
to the integral, and the expression can be rewritten in the following form
\eq
\langle F_i\rangle=\frac{\Omega_{d-1}}{2\Omega_d}\,
\int_0^1du\int\frac{d^d\bm{l}}{(2\pi)^d}\,\int_0^\pi d\theta\sin^{d-2}\theta
\left(\frac{\Phi^{(0)}_i(u;\theta)}{(\bm{l}^2-z^2)^2}
+\frac{\Phi^{(1)}_i(u;\theta)}{(\bm{l}^2-z^2)}\right)\, .
\en
The integral over $\bm{l}$ can be now evaluated. Since the ultraviolet divergences disappear in dimensional regularization, one may put $d=3$ afterwards, arriving at a two-dimensional integral representation for the $\rho$-meson form factors:
\eq
\langle F_i\rangle=\frac{1}{32\pi}\,\int_0^1du\int_0^\pi d\theta\frac{\sin\theta}{(-z^2)^{1/2}}\,\left(\Phi^{(0)}_i(u;\theta)+2z^2\Phi^{(1)}_i(u;\theta)\right)\, .
\en
Finally, the expression for the loop contribution can be multiplied by the pion form factor $F_\pi(k^2)$, in order to take into account all
high-order contributions in the pion-photon vertex.

\subsection{Convergence of the EFT expansion in dimensional regularization}

An astute reader may already notice a bottleneck in our arguments. Namely, in the
calculations we consistently ``pull out'' the momentum-dependent numerators from the
integrand, replacing them by their on-shell value. One could justify this procedure
by mentioning that
the difference between the two expressions vanishes in dimensional regularization
order by order in the low-energy expansion, since the integrand in this case represents a
low-energy polynomial. However, in the case of a resonance, the numerator is a singular
function of momenta (it namely becomes infinite on the real axis, exactly at the energy
where the phase shift becomes equal to the $90^\circ$), and the argument should be taken with a grain of salt. Stated
differently, the statement is even true at all orders, so the series diverges in the vicinity
of the resonance energy and hence, the problem requires further scrutiny.

Furthermore, note that exactly the same problem arises in the ordinary L\"uscher
equation, where the above solution is known to be correct. In fact, the L\"uscher equation can be
obtained without using the low-energy expansion and dimensional regularization at all.
The regular summation theorem holds, and in particular in the resonance region as
well, if the interaction range is of the natural size. This observation shows that the
above-mentioned controversy is of our own making, and hinges on the use of the
dimensional regularization and minimal subtraction in the derivation. If one uses, for example, the PDS
scheme~\cite{Kaplan:1998we} instead, the series converge at the resonance energy, and the final
answer looks exactly the same. One may now apply similar arguments to the calculation
of the resonance form factor, justifying the result obtained with the use of the shortcuts based on the standard
dimensional regularization.
 
\section{L\"uscher equation in the background field}
\label{sec:levels}

\subsection{Periodic electromagnetic field in a finite volume}
\label{sec:periodic}

The time-independent
periodic electromagnetic field on a cubic lattice of size $L$ is chosen
as~\cite{Lozano:2022kfz}
\eq\label{eq:A}
A^\mu(\bm{x},t)=\underline{A}^\mu\cos(\boldsymbol{\omega}\bm{x})\, ,\quad\quad\quad
\boldsymbol{\omega}=(0,0,\omega)\, ,\quad\omega=\frac{2\pi n}{L}\, ,\quad
n\in\mathbb{Z}\, .
\en
This allows one to extract the invariant form factors at the momentum transfer
$k^2=-\boldsymbol{\omega}^2$.
Furthermore, the freedom of choice of the constant vector $\underline{A}^\mu$ will be
used
to project out the invariant form factors. In order to understand this qualitatively,
let us recall that, at the lowest order, the interaction of the $\rho$-meson with the electromagnetic field is described by the scalar product of the form factor and the electromagnetic
field. Working in the Breit frame and taking $\alpha=0$, we get
\eq\label{eq:diag}
-\tilde\varepsilon_\mu(P,s)(P+Q)^0g^{\mu\nu}\varepsilon_\nu(Q,s')&=&2x\sqrt{s_R}
\left(\delta_{ss'}+2(x^2-1)\delta_{s3}\delta_{s'3}\right)\, ,
\nonumber\\[2mm]
-\tilde\varepsilon_\mu(P,s)(k^\mu g^{\nu 0}-k^\nu g^{\mu 0})\varepsilon_\nu(Q,s')&=&
-4x(x^2-1)\sqrt{s_R}\,\delta_{s3}\delta_{s'3}\, ,
\nonumber\\[2mm]
-\tilde\varepsilon_\mu(P,s)(P+Q)^0k^\mu k^\nu\varepsilon_\nu(Q,s')&=&
-2x^3(x^2-1)(s_R)^{3/2}\,\delta_{s3}\delta_{s'3}\, .
\en
Furthermore, taking $\alpha=3$ (vector potential
parallel to $\boldsymbol{\omega}$), one can easily ensure that the interaction with the
background electromagnetic field vanishes. Choosing however $\alpha=1$ or $\alpha=2$,
we get a single non-vanishing structure
\eq\label{eq:nondiag}
-\tilde\varepsilon_\mu(P,s)(k^\mu g^{\nu\alpha}-k^\nu g^{\mu\alpha})\varepsilon_\nu(Q,s')=
2x\sqrt{x^2-1}\sqrt{s_R}\,(\delta_{s'3}\delta_{s\alpha}-\delta_{s'\alpha}\delta_{s3})\, .
\en
In the simple case when the $\rho$-meson is stable, the scalar product of a form-factor and
the field $A^\mu$ gives merely the correction to the interaction Hamiltonian in a given kinematic configuration (fixed three-momenta $\bm{P},\bm{Q}$). Equations (\ref{eq:diag})
and (\ref{eq:nondiag}) show the modification of the diagonal $\delta H_{11}=\delta H_{22}$, $\delta H_{33}$ and non-diagonal $\delta H_{31}=\delta H_{32}$ matrix elements. The
finite-volume energy levels are obtained by diagonalizing the Hamiltonian. Three energy
levels are obtained after the diagonalization that enables to determine three invariant
form factors $G_1,G_2,G_3$.

In case the $\rho$-meson is unstable, the situation is more nuanced. The energy levels
are determined from the modified L\"uscher equation (see below). The counting of
the independent input/output parameters remains, however, unaffected.

\subsection{Modified L\"uscher equation}

The two-point function of a charged pion in a finite volume
is modified when the background electromagnetic field is turned on
\eq
S_+(p,q)=(2\pi)\delta(p_0-q_0)S_+(\bm{p},\bm{q};p_0)=i\int_L d^4x d^4y e^{ipx-iqy}\langle 0|T\phi_+(x)\phi_+^\dagger(y)|0\rangle_A\, .
\en
Note that in the periodic field~(\ref{eq:A}) the energy is conserved, as well as the components of momenta that are orthogonal to the vector $\boldsymbol{\omega}$.

The finite-volume spectrum is determined by the poles of the two-point function
of the interpolating field $\Psi_\mu(x)$ in the background field. When summing up the
bubble diagrams in this two-point function, one should distinguish between the interpolating field $\Psi_\mu(x)$ and the field $\Psi^{(1)}_\mu(x)$ that appears in the Lagrangian.
These two definitions differ by the off-shell terms. Note, however, that the ``off-shell''
field $\Psi_\mu(x)$ appears only at the beginning and the end of each diagram. Since
these ``endcaps'' cannot be responsible for the non-perturbative shift of the poles, one
may simply ignore the difference between the two fields and look for the poles of the
two-point function
\eq
D^L_{\mu\nu}(P,Q)&=&(2\pi)\delta(P_0-Q_0)D^L_{\mu\nu}(\bm{P},\bm{Q};P_0)
\nonumber\\[2mm]
&=&i\int_L d^4x d^4ye^{iPx-iQy}\langle 0|T\Psi^{(1)}_\mu(x)\Psi^{(1)}_\nu(y)|0\rangle_A\, .
\en
This two-point function obeys the equation
\eq\label{eq:LS}
D^L_{\mu\nu}(\bm{P},\bm{Q};P_0)&=&D^0_{\mu\nu}(\bm{P},\bm{Q};P_0)
\nonumber\\[2mm]
&+&\frac{1}{L^6}\sum_{\bm{P}'\bm{Q}'}
D^0_{\mu\sigma}(\bm{P},\bm{P}';P_0)\Pi^{\sigma\rho}(\bm{P}',\bm{Q}')
D^L_{\rho\nu}(\bm{Q}',\bm{Q};P_0)\, .
\en
Here, $D^0_{\mu\nu}(\bm{P},\bm{Q};P_0)$ is the two-point function in the absence of the four-point and five-point interactions in the Lagrangian, i.e., formally, $\sigma_1=0$ and
$g_1,g_2,g_3=0$:
\eq\label{eq:D0}
D^0_{\mu\nu}(\bm{P},\bm{Q};P_0)=\int\frac{dl_0}{2\pi i}\,\frac{1}{L^3}\sum_{\bm{l}}
\frac{\widetilde{W}_{P,\mu}f_Pf_Q\widetilde{W}_{Q,\nu}}{2W(W-(v\cdot l))}\,
S_+(\bm{P}-\bm{l},\bm{Q}-\bm{l};P_0-l_0)\, ,
\en
and $\Pi^{\mu\nu}(\bm{P},\bm{Q})$ plays the role of the ``self-energy''
\eq
\Pi^{\mu\nu}(\bm{P},\bm{Q})=\sigma_1g^{\mu\nu}L^3\delta^3_{\bm{P}\bm{Q}}
-L^3\delta^3_{(\bm{P}-\bm{Q}),\bm{k}}\hat G_b^{\alpha,\mu\nu}(P,Q)A_\alpha(\bm{k})\, ,
\en
where $\hat G_b^{\alpha,\mu\nu}(P,Q)$ is defined by Eq.~(\ref{eq:Gab}), and
$A_\alpha(\bm{k})$ is the Fourier-transform of the background field
\eq\label{eq:A-Fourier}
A_\alpha(\bm{k})=\int d^3\bm{x}e^{i\bm{k}\bm{x}}A_\alpha(\bm{x},0)
=\frac{1}{2}\,\underline{A}_\alpha L^3(\delta^3_{\bm{k}\boldsymbol{\omega}}+\delta^3_{\bm{k},-\boldsymbol{\omega}})\, .
\en
The equation (\ref{eq:LS}) defines the energy spectrum of two pions in the background
field. The derivation of the quantization condition is similar to the case with no background
field -- the position of the poles is determined by
\eq\label{eq:master}
\det(\mathscr{A})=0\,, \quad\quad
\mathscr{A}_{\mu\nu}=\Pi^{-1}_{\mu\nu}-D^0_{\mu\nu}\, .
\en
Note also
that all the quantities entering this expression are matrices both in the space of momenta
$\bm{P},\bm{Q}$, as well as Lorentz-indices $\mu,\nu$. In order to ease the notations,
the dependence on the momenta is suppressed.

\subsection{Projection on the irreps and the extraction of $g_1,g_2,g_3$ from data}

According to the discussion in Sect.~\ref{sec:periodic}, it suffices to consider two different
background field configurations: the timelike configuration with $\underline{A}^\alpha=g^{\alpha 0}\underline{A}$
and the transverse one, $\underline{A}^\alpha=g^{\alpha 1}\underline{A}$.
We further define\footnote{Here, $P^2,Q^2$ are both on the real axis,
and the polarization vector $\tilde\varepsilon_\mu(P,s)$ coincides
with $\varepsilon_\mu^*(P,s)$ (the same for $\tilde\varepsilon_\mu(Q,s')$).}
\eq
\tilde\varepsilon^{\mu}(P,\lambda)=(iw^\mu,\tilde\varepsilon^{\mu}(P,s))\, ,\quad\quad
\lambda=(0,s)\, ,
\nonumber\\[2mm]
\varepsilon^{\nu}(Q,\lambda')=(iu^\nu,\varepsilon^{\nu}(Q,s'))\, ,\quad\quad
\lambda'=(0,s')\, .
\en
These vectors satisfy the condition
\eq
\sum_\lambda \varepsilon^\mu(P,\lambda)\tilde\varepsilon^{\nu}(P,\lambda)
=
\sum_\lambda \varepsilon^\mu(Q,\lambda)\tilde\varepsilon^{\nu}(Q,\lambda)
=-g^{\mu\nu}\, .
\en
Projecting the quantization condition in the basis of the indices gives
\eq
\mathscr{A}_{\lambda,\lambda'}=\Pi^{-1}_{\lambda,\lambda'}-D^0_{\lambda,\lambda'}
=\tilde\varepsilon^{\mu}(P,\lambda)\biggl(
\Pi^{-1}_{\mu\nu}(\bm{P},\bm{Q};P_0)-D^0_{\mu\nu}(\bm{P},\bm{Q};P_0)\biggr)
\varepsilon^{\nu}(Q,\lambda')\, .
\en
In the case of the timelike background field configuration, it can be directly checked that the nonzero matrix
elements of $\Pi^{-1}_{\lambda,\lambda'}$ up to $O(e^2)$ are given by
\eq
\Pi^{-1}_{00}&=&L^3\delta^3_{\bm{P}\bm{Q}}(2x^2-1)
+L^3\delta^3_{\bm{P}-\bm{Q},\bm{k}}\underline{A}(\bm{k})
2eP_0k^2\biggl(-g_1(2x^2-1)-g_2\frac{1}{2P^2}+g_3\frac{k^2}{2P^2}\biggr)\, ,
\nonumber\\[2mm]
-i\Pi^{-1}_{03}&=&i\Pi^{-1}_{30}
=L^3\delta^3_{\bm{P}\bm{Q}}2x\frac{|\bm{P}|}{\sqrt{P^2}}
-L^3\delta^3_{\bm{P}-\bm{Q},\bm{k}}\underline{A}(\bm{k})
\nonumber\\[2mm]
&&\quad\quad\quad \times\,\, 2x\frac{|\bm{P}|}{\sqrt{P^2}}\,2eP_0\biggl(-g_1k^2+\frac{1}{2}\,g_2\biggl(1-\frac{k^2}{4x^2P^2}\biggr)-g_3k^2\biggr)\, ,
\nonumber\\[2mm]
\Pi^{-1}_{11}&=&\Pi^{-1}_{22}=
L^3\delta^3_{\bm{P}\bm{Q}}+L^3\delta^3_{\bm{P}-\bm{Q},\bm{k}}\underline{A}(\bm{k})2eP_0g_1k^2\, ,
\nonumber\\[2mm]
\Pi^{-1}_{33}&=&\Pi^{-1}_{11}+
L^3\delta^3_{\bm{P}\bm{Q}}\frac{\bm{P}^2}{P^2}
+L^3\delta^3_{\bm{P}-\bm{Q},\bm{k}}\underline{A}(\bm{k})
4eP_0\bm{P}^2\biggl(g_1\frac{k^2}{2P^2}-g_2\frac{1}{P^2}-4g_3x^2\biggr)\, .
\en
One can also directly verify that the matrix $D^0_{\lambda,\lambda'}$ has exactly the same
symmetry structure as $\Pi^{-1}_{\lambda,\lambda'}$. Hence, both the matrices have block-diagonal form and the quantization condition factorizes, corresponding
to $\mathscr{A}=\mathscr{A}_1\times\mathscr{A}_2$, where $\mathscr{A}_1$ and
$\mathscr{A}_2$ are $2\times 2$ matrices composed of the 0th,3rd rows/columns and
the 1st,2nd rows/columns of the $4\times 4$ matrix $\mathscr{A}$. In group-theoretical
language, this corresponds to the projection onto the irreps $A_1$ and $E$ of the little group $C_{4v}$ of the octahedral group~\cite{Gockeler:2012yj} (in the case of irrep $E$, the corresponding matrix $\mathscr{A}_2$ is diagonal and hence the dimensionality of the problem can be further reduced).

In the case of the transverse background field configuration, $C_{4v}$ is no more the symmetry of the problem and $\mathscr{A}$ is not block-diagonal anymore. The nonzero entries of the matrix $\Pi^{-1}_{\lambda,\lambda'}$ are
\eq
\Pi^{-1}_{00}&=&
L^3\delta^3_{\bm{P}\bm{Q}}(2x^2-1)\, ,
\nonumber\\[2mm]
-i\Pi^{-1}_{03}&=&i\Pi^{-1}_{30}
=L^3\delta^3_{\bm{P}\bm{Q}}2x\frac{|\bm{P}|}{\sqrt{P^2}}\, ,
\nonumber\\[2mm]
-i\Pi^{-1}_{01}&=&-i\Pi^{-1}_{10}
=-L^3\delta^3_{\bm{P}-\bm{Q},\bm{k}}\underline{A}(\bm{k})
g_2e\frac{k^2}{2\sqrt{P^2}}\, ,
\nonumber\\[2mm]
\Pi^{-1}_{13}&=&-\Pi^{-1}_{31}
=-L^3\delta^3_{\bm{P}-\bm{Q},\bm{k}}\underline{A}(\bm{k})
2g_2ex|\bm{P}|\, ,
\nonumber\\[2mm]
\Pi^{-1}_{11}&=&\Pi^{-1}_{22}
=L^3\delta^3_{\bm{P}\bm{Q}}\, ,
\nonumber\\[2mm]
\Pi^{-1}_{33}&=&\Pi^{-1}_{11}
+L^3\delta^3_{\bm{P}\bm{Q}}\frac{\bm{P}^2}{P^2}\, .
\en
Finally, note that we are interested in the structure of the energy levels at the first
order in $e$. As demonstrated in Ref.~\cite{Lozano:2022kfz}, at this order each entry
of the matrix $\mathscr{A}_{\lambda\lambda'}$ has two components corresponding to
$|\bm{P}|,|\bm{Q}|=\pm|\bm{k}|/2$ in the Breit frame (the transverse component of
the three-vectors $\bm{P},\bm{Q}$ is conserved and can be set to zero from the
beginning). It has also been shown that other entries to this matrix, corresponding to
$|\bm{P}|,|\bm{Q}|=\pm n|\bm{k}|/2$, $n>1$,
contribute at $O(e^n)$ and can be safely neglected.
With this, we finally arrive at the L\"uscher equation in the background field that
determines finite-volume energy levels at $O(e)$. These are the three equations in two
different background field configurations that allow one to fit the couplings $g_1,g_2,g_3$
to lattice data.

\subsection{Calculation of the loop in the background field}

According to Ref.~\cite{Lozano:2022kfz}, there are two ways of calculating the loop
given in Eq.~(\ref{eq:D0}),
the perturbative and the non-perturbative one. The perturbative calculation is
straightforward. The propagator of the charged pion, which enters Eq.~(\ref{eq:D0}),
obeys the following equation
(recall that we work in the Breit frame with $v^\mu=(1,\bm{0})$)
\eq\label{eq:Splus}
S_+(\bm{p},\bm{q};p_0)=L^3\delta^3_{\bm{p}\bm{q}}S_+^0(\bm{p},p_0)
+\frac{1}{L^3}\,\sum_{\bm{k}}S_+^0(\bm{p},p_0)A_\alpha(\bm{p}-\bm{q})
N^\alpha(p,q)S_+(\bm{k},\bm{q};p_0)\, .
\en
Here,
\eq
S_+^0(\bm{p},p_0)=\frac{1}{2W(p)(W(p)-p_0)}
\en
Furthermore, the Fourier-transform of the background field is given by
Eq.~(\ref{eq:A-Fourier}), and the vertex $N^\alpha$ is given in Eq.~(\ref{eq:Nmu})
(in this expression one should take $p_0=q_0$). From Eq.~(\ref{eq:Splus}) at $O(e)$
one gets
\eq
S_+(\bm{p},\bm{q};p_0)=L^3\delta^3_{\bm{p}\bm{q}}S_+^0(\bm{p},p_0)
+S_+^0(\bm{p},p_0)A_\alpha(\bm{p}-\bm{k})
N^\alpha(p,q)S_+^0(\bm{q};p_0)+O(e^2)\, .
\en
Substituting this expression into Eq.~(\ref{eq:D0}), one gets the perturbative expression
for the loop that can be evaluated straightforwardly. In particular, at $O(e)$ the calculation
can be done by a method identical to the one used above for the calculation of the
infinite-volume form factor. The only difference consists in replacing the three-dimensional
integral by the finite-volume sums. We do not display the result here and refer to
Ref.~\cite{Lozano:2022kfz}, where similar calculations have been carried out in detail.

As pointed out in Ref.~\cite{Lozano:2022kfz}, the perturbative calculation of the
pion loop, corresponding to the ``L\"uscher zeta-function in the presence of the
background field'' yields very accurate results {\em almost} everywhere on the real
energy axis, except in the immediate vicinity of the singularities of $S_+^0(\bm{p},p_0)$.
Therefore, the perturbative expression should be used with caution. For example, as
shown in Ref.~\cite{Lozano:2022kfz}, the analysis done in the perturbative framework may lead to a peculiar behavior of the finite-volume energy levels, namely, to the merger
and disappearance of the energy levels, the emergence of the spurious solutions and
so on. Alternatively, one could use the non-perturbative expression for the loop function,
and ensure that the analysis of data is devoid of such unpleasant surprises.

Let us first start with the case of the Green function in the timelike background field.
In this case, $N_0(p,q)=e(W(p)+W(q))$. The spectral representation of
the Green function defined by the integral equation (\ref{eq:Splus}) can be
written down in terms of the eigenvalues and eigenvectors of the following equation
\eq\label{eq:eigen}
2W(p)(W(p)-E_n)\psi_n(\bm{p})
+\frac{e}{L^3}\,\sum_{\bm{q}}A_0(\bm{p}-\bm{q})(W(p)+W(q))\psi_n(\bm{q})=0\, ,
\en
where the eigenvectors are normalized as
\eq
\frac{1}{L^3}\sum_{\bm{p}}\frac{1}{2W(p)}\,\psi_n(\bm{p})\psi_m(\bm{p})=\delta_{mn}\, ,
\en
and the Green function is given by
\eq
S_+(\bm{p},\bm{q};p_0)=\sum_m\frac{\psi_n(\bm{p})\psi_n(\bm{q})}{\sqrt{2W(p)}(E_n-p_0)\sqrt{2W(q)}}\, .
\en
Equation~(\ref{eq:eigen}) does not have a standard form of the Mathieu equation, but
can be reduced to it at $O(e)$. To this end, we define
\eq
\psi_n(\bm{p})=\phi_n(\bm{p})+\frac{e}{2W(p)}\,\frac{1}{L^3}\,\sum_{\bm{q}}
A_0(\bm{p}-\bm{q})\phi_n(\bm{q})\, .
\en
Dropping the $O(e^2)$ terms, Eq.~(\ref{eq:eigen}) transforms into
\eq
(W(p)-E_n)\phi_n(\bm{p})+\frac{e}{L^3}\,\sum_{\bm{q}}
A_0(\bm{p}-\bm{q})\phi_n(\bm{q})=0\, ,
\en
or, dropping again $O(e^2)$ terms,
\eq\label{eq:squaring}
(W^2(p)-E_n^2)\phi_n(\bm{p})+\frac{2eE_n}{L^3}\,\sum_{\bm{q}}
A_0(\bm{p}-\bm{q})\phi_n(\bm{q})=0\, .
\en
This equation has the form of the Mathieu equation with the interaction strength depending
on the eigenvalue $E_n$. Hence, the eigenvalues (and eigenvectors)
can be calculated perturbatively
using the well-known expansion of the eigenvalues of the Mathieu equation in the small
parameter $e$. Furthermore, since
corrections to the most of the eigenvalues start at $O(e^2)$,
the perturbative approach will be needed for selected eigenvalues only.

Next, consider the case of the transverse background field configuration. Introducing
the notation $A(\bm{p},\bm{q})=-(p+q)_1A_1(\bm{p}-\bm{q})$, we arrive at the
equation
\eq
2W(p)(W(p)-E_n)\psi_n(\bm{p})
+\frac{2e}{L^3}\,\sum_{\bm{q}}A(\bm{p},\bm{q})\left(
1-\frac{E_n}{W(p)+W(q)}\right)\psi_n(\bm{q})=0\, .
\en
This equation can again be reduced to the Mathieu equation by the transformations of
the wave function. In particular, rescaling $\psi_n(\bm{p})=\dfrac{1}{\sqrt{2W(p)}} \,\phi^{(1)}_n(\bm{p})$, one gets
\eq
(W(p)-E_n)\phi^{(1)}_n(\bm{p})
+\frac{e}{2E_nL^3}\,\sum_{\bm{q}}A(\bm{p},\bm{q})B(\bm{p},\bm{q};E_n)
\phi_n^{(1)}(\bm{q})=0\, ,
\en 
where
\eq
B(\bm{p},\bm{q};E_n)
&=&\left(1+\frac{E_n-W(p)}{\sqrt{W(p)}(\sqrt{E_n}+\sqrt{W(p)})}\right)
\left(1-\frac{E_n-W(p)}{W(p)+W(q)}-\frac{E_n-W(q)}{W(p)+W(q)}\right)
\nonumber\\[2mm]
&\times&\left(1+\frac{E_n-W(q)}{\sqrt{W(q)}(\sqrt{E_n}+\sqrt{W(q)})}\right)
\nonumber\\[2mm]
&=&1+\alpha(\bm{p},\bm{q};E_n)(E_n-W(p))+\beta(\bm{p},\bm{q};E_n)(E_n-W(q))
\, .
\en
Note that the exact form of the coefficients $\alpha,\beta$ is not important.
The eigenvalue equation can be then rewritten as
\eq
&&(W(p)-E_n)\frac{1}{L^3}\,\sum_{\bm{q}}\left(L^3\delta_{\bm{p}\bm{q}}
-\frac{e}{2E_n}\,\alpha(\bm{p},\bm{q};E_n)A(\bm{p},\bm{q})\right)
\phi^{(1)}_n(\bm{q})
\nonumber\\[2mm]
&+&\frac{e}{2E_nL^3}\,\sum_{\bm{q}}A(\bm{p},\bm{q})\left(1
+\beta(\bm{p},\bm{q};E_n)(E_n-W(q))\right)
\phi_n^{(1)}(\bm{q})=0\, .
\en
Defining now
\eq
\phi^{(1)}_n(\bm{p})
=\frac{1}{L^3}\,\sum_{\bm{q}}\left(L^3\delta_{\bm{p}\bm{q}}
+\frac{e}{2E_n}\,\alpha(\bm{p},\bm{q};E_n)A(\bm{p},\bm{q})\right)
\phi^{(2)}_n(\bm{p})\, ,
\en
it can be seen that the off-shell term multiplied by $\alpha$ disappears, and the equation takes
the form
\eq
&&\frac{1}{L^3}\,\sum_{\bm{q}}\left(L^3\delta_{\bm{p}\bm{q}}
-\frac{e}{2E_n}\,\beta(\bm{p},\bm{q};E_n)A(\bm{p},\bm{q})\right)
(W(q)-E_n)\phi^{(2)}_n(\bm{q})
\nonumber\\[2mm]
&+&\frac{e}{2E_nL^3}\,\sum_{\bm{q}}A(\bm{p},\bm{q})
\phi^{(2)}_n(\bm{q})=0\, .
\en
It is now evident that the off-shell term proportional to $\beta$ can be also dropped
at $O(e)$, leading to a much simpler expression
\eq\label{eq:2}
(W(p)-E_n)\phi^{(2)}_n(\bm{q})
+\frac{e}{2E_nL^3}\,\sum_{\bm{q}}A(\bm{p},\bm{q})
\phi^{(2)}_n(\bm{q})=0\, .
\en
The spectral representation for the Green function of this equation can be written down
straightforwardly
\eq\label{eq:S2}
S^{(2)}(\bm{p},\bm{q};p_0)=\sum_n\frac{\phi^{(2)}_n(\bm{p})\phi^{(2)}_n(\bm{q})}
{(E_n-p_0)}\, ,
\en
where the eigenfunctions are normalized according to
\eq
\frac{1}{L^3}\,\sum_{\bm{p}}\phi^{(2)}_n(\bm{p})\phi^{(2)}_m(\bm{p})
=\delta_{nm}\, .
\en
Note that Eq.~(\ref{eq:2}) still does not have the standard form of the Mathieu equation.
Squaring the operator acting on $\phi^{(2)}_n(\bm{p})$ in analogy to Eq.~(\ref{eq:squaring}), we finally arrive at
\eq
(W^2(p)-E_n^2)\phi^{(2)}_n(\bm{q})
+\frac{e}{L^3}\,\sum_{\bm{q}}A(\bm{p},\bm{q})
\phi^{(2)}_n(\bm{q})=0\, .
\en
Note also that in the above equation interaction strength, unlike Eq.~(\ref{eq:squaring}),
does not depend on the eigenvalue $E_n$, but depends on the transverse momentum
$(p+q)_1$, which enters in the definition of $A(\bm{p},\bm{q})$. Finally, the expression
for the Green function $S_+(\bm{p},\bm{q};p_0)$ we are looking for can be obtained
from Eq.~(\ref{eq:S2}) by applying the reverse chain of transformations on the wave functions entering this spectral representation. 

An explicit calculation of the loop with the exact charged pion propagator is
a very challenging task. As discussed in Ref.~\cite{Lozano:2022kfz}, a substantial
simplification is achieved if only the {\em numerator} of the spectral representation
(which is a non-singular function) is expanded up to and including $O(e)$, whereas
the denominator stays intact. We relegate this rather straightforward
but still voluminous calculation to some future publication and refer an
interested reader to Ref.~\cite{Lozano:2022kfz}, where the method is discussed
in great detail.

\section{Implementation, comparison to the alternative approach}
\subsection{Lattice calculation}

To compute the resonance form factors, all that is needed is the knowledge of the energy
levels in the presence of the background field close to the resonance.
The lattice Lagrangian is modified by
\eq\label{eq:L_lambda}
\mathscr{L}(x) \rightarrow \mathscr{L}(x) + \lambda Z_V \sum_qe_q \bar q(x) \gamma_\mu A_\mu(x) q(x) \,, 
\en
where $A_\mu$ is given by Eq.~(\ref{eq:A}),
$e_q$ is the electric charge of the quark with flavor $q$ and $Z_V$ denotes the
renormalization constant of the local vector current. One could also use the
conserved vector current in Eq.~(\ref{eq:L_lambda}), for which $Z_V=1$ (preferred).
The gauge fields need to be simulated for several values of $\lambda$, in
order to perform the extrapolation to $\lambda = 0$, and for a range of momentum
transfers $\boldsymbol{\omega}$. The standard procedure for computing the energy
levels is GEVP. The calculation is done in the Breit frame. To simplify the calculation,
one may however restrict oneself to $\bar{q} \gamma_\mu q$ sources and
sinks~\cite{Gockeler:2008kc}. There are three distinct sets of energy levels,
corresponding to the different irreps and the different orientations of the background field,
from which three independent form factors $G_1$, $G_2$ and $G_3$ can be extracted.

At the next step, the energy levels $E_n(\lambda,k^2)$ are substituted into the
L\"uscher equation~(\ref{eq:master}), which is fitted to the couplings
$g_1(k^2)$, $g_2(k^2)$ and $g_3(k^2)$ with $k^2 = -\boldsymbol{\omega}^2$
(the fit is carried out for different values of $k^2$ separately, that effectively results
in the $k^2$-dependence of these couplings).
Note that, by definition, neither these
couplings nor the phase shift, which is encoded in the vertex function $f_\ell$ through
Eq.~(\ref{eq:cot}), depend on $\lambda$. Using the explicit relations between
$g_1$, $g_2$, $g_3$ and $G_1$, $G_2$, $G_3$ derived above, we finally obtain the
form factors $G_1(k^2)$, $G_2(k^2)$ and $G_3(k^2)$ we are looking for. Note that
the triangle diagram is computed separately and the final result is multiplied
by the form factor of the charged pion, $F_\pi(k^2)$ (in order to ease the
notations, we set $F_\pi(k^2)=1$ in all expressions in the text).

Alternatively, one may compute the position of the resonance pole $W(\lambda,k^2)$
in the complex plane. This can be achieved by first determining the couplings
$g_1(k^2), g_2(k^2)$ and $g_3(k^2)$ and then solving the infinite-volume counterpart
of the same L\"uscher equation in different irreps and for the different background field.
The energy $E$ and the width $\Gamma$ of a resonance are related to $W(\lambda,k^2)$
by $W = E - i\Gamma/2$. Now, by differentiating $W(\lambda,k^2)$ with respect to $\lambda$,
one arrives exactly at the same $G_1(k^2)$, $G_2(k^2)$ and $G_3(k^2)$, in accord with
the Feynman-Hellmann theorem. The proof follows the lines of~\cite{Lozano:2022kfz}. 
However, this method perhaps lends itself to a direct calculation outlined above.\footnote{In the calculation
of a resonance form factor at {\em zero momentum transfer} -- notably, the resonance
$\sigma$ terms -- it is possible to avoid this detour and to extract the
physical form factor directly with the use of the Feynman-Hellmann theorem~\cite{RuizdeElvira:2017aet}.
Unfortunately, the same trick does not work for arbitrary kinematics.}

\subsection{Comparison to the alternative approach}
As already mentioned in the introduction, in the alternative approach which is
described in Refs.~\cite{Hoja:2010fm,Bernard:2012bi,Baroni:2018iau,Briceno:2019nns,Briceno:2020vgp,Briceno:2020xxs,Moscoso:2026wmz}, one starts with the evaluation of the three-point function on the lattice. The method of doing so is pretty standard and the finite-volume matrix element is a well-defined quantity. At the next stage
one analytically removes all factors that exhibit power-law dependence on the box size $L$  from this quantity. These are: a) infinite bubble chains in the external lines that disappear after multiplying
by a Lellouch-L\"uscher factor for each line, and b) the triangle diagram. Only the short-range contribution (a counterpart of the contact contribution in the present approach)
is left, which receives exponentially suppressed corrections
in a finite volume. Performing the infinite-volume limit in this contribution, one then
re-introduces the infinite-volume triangle diagram by hand.

In difference to this, in the present approach there is no need to calculate the matrix element of the current on the lattice. Instead, one calculates the finite-volume
energy spectrum of a system in an external periodic electromagnetic field. The short-range contribution to the form-factor can be directly extracted from the modified L\"uscher equation in the external field. The last step -- adding the infinite-volume triangle diagram --
is identical in both approaches.

A decisive argument in favor of one or another approach should be provided by its relative simplicity and convenience. In our approach, one calculates energy spectrum of a system instead of
the matrix element that is a big advantage. On the other hand, calculations in the
external field require generation of new lattice configurations (this problem can be alleviated in a $SU(3)$-symmetric QCD where electroquenching is justified). 
It is therefore very difficult to make a judgment at the present stage -- clearly, the
final word will be said by lattice practitioners.

\section{Numerical implementation: contact contributions to the form factors in NREFT}
\label{sec:numerics}

\begin{figure}[t]
\begin{center}
\includegraphics[width=8.cm]{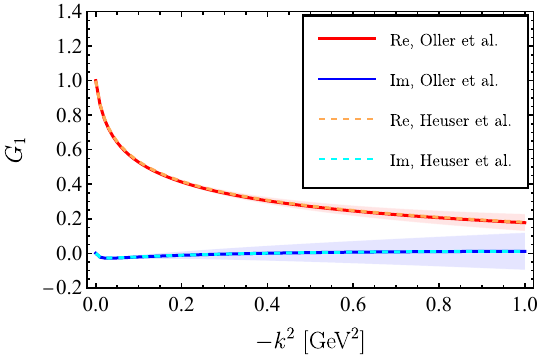}
\vspace*{0.5em}

\includegraphics[width=8.cm]{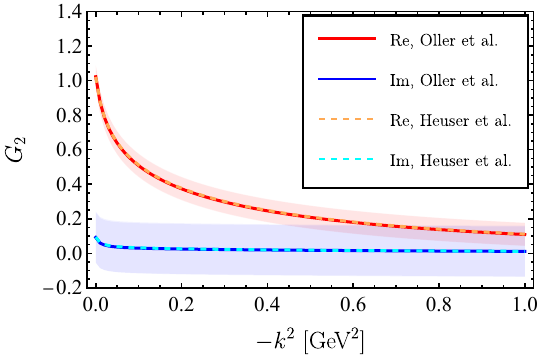}
\vspace*{0.5em}

\includegraphics[width=8.cm]{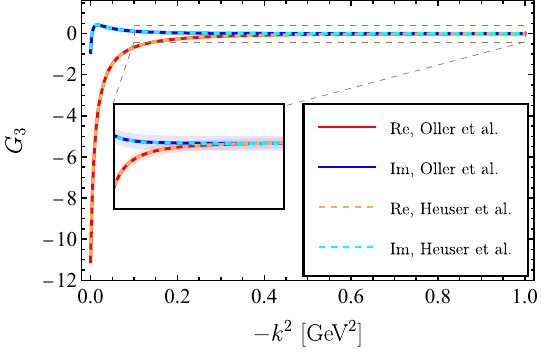}
\caption{Invariant form factors $G_1(k^2),G_2(k^2),G_3(k^2)$ up to
$-k^2\le (1\,\mbox{GeV})^2$. Two different phenomenological parameterizations of the $\pi\pi$ amplitude are used. The shaded areas correspond to the crude
estimate of the low-energy couplings $g_1,g_2,g_3$ at the next-to-leading order
in ChPT. For better visibility of the bands, the inset on the third plot shows an enlarged region for
$0.1\,\mbox{GeV}^2\leq -k^2\leq 1\,\mbox{GeV}^2$.}
\label{fig:G123}
\end{center}
\end{figure}

\begin{figure}[t]
\begin{center}
\includegraphics[width=8.cm]{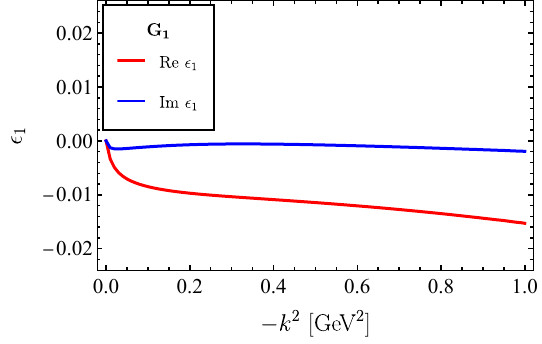}
\vspace*{0.5em}

\includegraphics[width=8.cm]{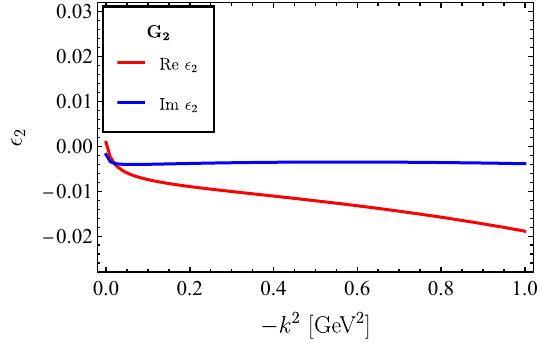}
\vspace*{0.5em}

\includegraphics[width=8.cm]{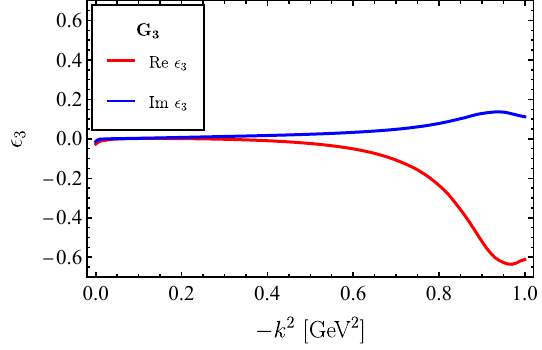}
\caption{The real and imaginary parts of the function $\epsilon_i(k^2)$ (a relative
error resulting from the use of two different phenomenological parameterizations),
which is given
in Eq.~(\ref{eq:epsilon}). As seen, in a rather large interval, the magnitude of this function amounts up to a few percent.}
\label{fig:G123-epsilon}
\end{center}
\end{figure}

One of the objectives of the present paper is to estimate the relative
weight of the contact
contributions to the form factors of the $\rho$-meson. We remind the reader that the other
contribution to the form factor, given by the triangle diagram, is written down in terms
of the $\pi\pi$ scattering amplitude that can be determined independently from the
calculation of the form factors on the lattice.

In order to achieve this goal, we numerically calculate all three invariant form
factors in Eq.~(\ref{eq:final-G}), using a crude
estimate of the low-energy
couplings $g_1,g_2,g_3$ from ChPT and the
phenomenological parameterizations for the
$\pi\pi$ amplitude from Refs.~\cite{Heuser:2024biq,Oller:1998hw}, see also
Appendix~\ref{app:chanturia}. 

The two parameterizations that are used in the calculations agree very well on the real
axis up to $1\,\mbox{GeV}$ in the center-of-mass energy.\footnote{The parameterization
suggested in Ref.~\cite{Schenk:1991xe} is not very convenient for the numerical implementation, and we do not use it here.} The calculation of the form factor involves, however,
the analytic continuation into the complex plane. Is the final result sensitive to small
differences in the input on the real axis? If it happens to be so, one would have to
conclude that the proposed method is numerically unstable. We address this issue explicitly and, moreover, scan all values of photon virtuality
$k^2$ in the interval $-1\,\mbox{GeV}^2\leq k^2\leq 0$ in order to identify the potentially dangerous regions.

Furthermore, a very crude estimate of the low-energy couplings, carried out at the
next-to-leading order in ChPT in Appendix~\ref{app:ChPT}, gives
$g_2=2.8\cdot 10^{-2}M_\pi^{-2}$. The estimates for $g_1$, $g_3$ are given in
Eq.~(\ref{eq:g_13}). It should be noted that these are only rough estimates, since they are calculated at the threshold,
and include the contributions only up to the tree level at $O(p^4)$. See Appendix~\ref{app:ChPT} for more details.
Calculating the contact contribution
from these couplings, we sweep all values from $-g_i$ to $+g_i$. Again, we stress
that this is by no means a conservative estimate of the uncertainty, and the absolute
values of the couplings $g_i$ could easily be outside this interval (for example, owing to
the chiral logs in ChPT).

After these preliminary remarks, we present the results of
numerical calculations of all three invariant form factors in Fig.~\ref{fig:G123}.
The bands in these figures
correspond to the variation of the effective couplings from $-g_i$ to $+g_i$, whereas the solid lines are central values at $g_i=0$. As seen
from the figure, the contact contribution is substantial in all three form factors
(especially in the imaginary part) that justifies the interest to their study
on the lattice. Also, the difference between the results obtained with the use of
two different parameterizations is not seen by bare eye. In other words, the final
result is very stable with respect to minor perturbations on the real axis.

Furthermore, small differences between two parameterizations can be parameterized by introducing the quantity
\eq\label{eq:epsilon}
\epsilon_i(k^2)=\frac{2(G_i(k^2)-G_i'(k^2))}{|G_i(k^2)|+|G_i'(k^2)|}\, ,\quad\quad
i=1,2,3\, ,
\en
where $G_i(k^2)$ and $G_i'(k^2)$ are calculated with the use of the parameterizations from Ref.~\cite{Oller:1998hw} and
Ref.~\cite{Heuser:2024biq},
respectively. The plots of the quantity $\epsilon_i(k^2)$ are given in the Fig.~\ref{fig:G123-epsilon}. Almost in the whole interval of the variable
$k^2$, the difference amounts to a few percent only. Blowing up of $\epsilon_3(k^2)$ for larger negative values of $k^2$ can be attributed to the fact that the form factor
goes to zero very rapidly in this region (i.e., the denominator in Eq.~(\ref{eq:epsilon})
becomes very small). To summarize, one may conclude that the obtained results
are very robust and stable with respect to small variations in the input parameterization.

Furthermore, from Fig.~\ref{fig:G123}
it can be seen that the form factors $G_1(k^2),G_2(k^2)$ are of natural
size, whereas $G_3(k^2)$ turns out large (but still finite) near $k^2=0$.
We have checked that the large contribution comes
from a single term in the expression containing the term $(x^2-1)^{-1}$,
see Eq.~(\ref{eq:barF}). For very small $|k^2|$, the first and second derivatives
of the $\pi\pi$ amplitude in the vicinity of the pole contribute to this term. The reason
why these derivatives should be large is obvious: the tangent of the phase shift which
determines the $\rho\pi\pi$ vertices $f_P(q^2),f_Q(q^2)$, blows up and becomes infinite
on the real axis very close
to the (complex) pole. Consequently, each derivative of these functions in the
variable $q^2$,
evaluated at the pole position, contains the small scale factor $\Gamma_R^2$ in
the denominator ($\Gamma_R$ denotes the $\rho$-meson width), and the higher-order
derivatives can become large. Note that this problem does not occur in the form factors 
$G_1(k^2),G_2(k^2)$ which are free of the above-mentioned kinematic
singularities.\footnote{Note that in $G_3(k^2)$ a ``milder'' singularity proportional
to $(x^2-1)^{-1/2}$ is also present. Its contribution, however, is small as compared to
the term with the leading singularity.} Needless to say, this is a non-perturbative
effect, since the singularity in the function $f(q^2)$ cannot be obtained in any finite
order.

Last but not least, the function $f(q^2)$ here has been determined from the fit to the experimental data. On the lattice, one usually has ensembles corresponding to non-physical quark masses. Then, $f(q^2)$ should be fitted to the scattering phase shifts calculated for the same quark mass by using the L\"uscher formula. If the quark masses are not very far from the physical ones, we expect that the same analytic form will suffice -- only the parameters should be re-fit.

We would like to put particular emphasis on two results that were obtained from the calculations in the infinite volume:
\begin{enumerate}
\item
The quantity $G_2(0)$ determines the magnetic moment of the
$\rho$-meson in proper units. The central value for this quantity
$G_2(0)=1.05$, obtained from the
triangle diagram, is very robust and almost parameter-free (only P-wave $\pi\pi$ phase
enters here which is very well known in the low-energy region). The uncertainty at leading order comes from the coupling $g_2$. Even increasing the NLO ChPT estimate of this
constant, given in Eq.~(\ref{eq:matching-ChPT}) by a factor 5 in order to account for the (hypothetical) chiral logarithm, we arrive at the result $G_2(0)=1.05\pm 0.3$. On the experimental side,
there exists no direct way to determine the magnetic dipole moment of the
$\rho$-meson. An indirect method implies the use of the vector meson dominance
models in the analysis of BABAR data~\cite{GarciaGudino:2013alv,GarciaGudino:2015ocw,Rojas:2024tmn}.
The calculation of the magnetic moment was also addressed
within various phenomenological approaches
and models~\cite{Aliev:2004uj,Braguta:2004kx,Choi:2004ww,Bhagwat:2006pu,Carrillo-Serrano:2015uca}, as
well as in ChPT in the presence of the vector mesons~\cite{Djukanovic:2013mka}.
The calculation of this quantity on the lattice involves heavier quark masses,
for which the $\rho$-meson is stable~\cite{Lee:2008qf,QCDSF:2008tjq}. As a rule, all these
methods end up with a value which is considerably higher that the one obtained in our work. Taking into account the fact that there is no substantial source of systematic uncertainty is left in our approach, it would be extremely interesting to get a direct check of this result on the lattice. In this connection, note that the the minimal substitution
does not uniquely determine the magnetic moment of the spin-1 particle
in difference to the spin-$\frac{1}{2}$ particles~\cite{TDLee}. There are reasons to
believe that a value close to 2 is a preferred choice for the massive weak vector
bosons but,
to the best of our knowledge, no similar reasons exist in case of the $\rho$-meson.
If this value indeed turns out to be close to 2, one has to conclude that the ChPT
order-of-magnitude estimate for the coupling $g_2$ fails completely.

\item
The quadrupole moment which is related to $G_3(0)$ turns out
extremely large, albeit finite (for example, $|\mbox{Re}\,G_3(0)|\simeq 10$ in proper units). This result
is also very general and comes almost completely from the triangle diagram alone.
Namely, this large contribution stems from the presence of the
kinematical singularity in the form factor which
results in the derivatives of the phase shift. So, formally, the result would diverge
for an infinitely narrow resonance and should be very large in case of a
narrow resonance. It is also seen that the result is non-perturbative in nature, since
no resonance emerges in the calculations at finite order. In addition, the curvature
of the form factor near $k^2=0$ turns out very large for the same reason.
Again, an independent check of this (very robust) prediction on the lattice would be of a major interest.
\end{enumerate}

\section{Conclusions}
\label{sec:concl}

\begin{itemize}
\item[i)]
In this paper we have formulated a manifestly Lorentz-invariant framework
for the extraction of the $\rho$-meson electromagnetic form factor on the lattice.
The framework, which is derived on the basis of NREFT, circumvents the problem
of the irregular behavior of the triangle diagram in a finite volume. Here, one extracts
the contact contribution to the form factor characterized by the low-energy
couplings $g_1,g_2,g_3$ (at the leading order). These couplings receive only
exponentially suppressed corrections in a finite volume. The infinite-volume form factor
is then {\em calculated} in NREFT, using $g_1,g_2,g_3$ as an input.

\item[ii)]
The background field method (the Feynman-Hellman method) is used on the lattice
for the extraction of the couplings $g_1,g_2,g_3$. Namely, these are determined from
the fit to the energy shift in the
background field, obtained from the two-point function of the resonance field. Measuring
matrix elements (three-point functions) is superfluous in this method.

\item[iii)]
The $\pi\pi$ scattering amplitude provides an important input and can be determined
independently from the form factor. The data, obtained on the real axis by using
the L\"uscher equation should be continued to the complex $\rho$-meson pole with the help of
explicit parameterizations. We have checked that the numerical procedure
used here is very stable with respect to the small variations of the input on the real axis.

\item[iv)] Performing the matching to the four-pion-photon amplitude at threshold
in ChPT at next-to-leading order and using dimensional arguments in addition,
we were able to obtain a very crude estimates of the couplings $g_1,g_2,g_3$.
Evaluating the form factors $G_1(k^2),G_2(k^2),G_3(k^2)$ with this input, we
see that the triangle graph dominates but the contact contribution is still
substantial, justifying the interest to the study of the problem on the lattice.
It should be stressed that this result was obtained with the systematic uncertainty
in $g_i$ grossly underestimated. In other words, the effect can eventually
turn even larger that renders the lattice calculation of the $\rho$-meson form
factor even more intriguing.

\item[v)]
We have obtained a very robust and largely model-independent
predictions for the magnetic and quadrupole moments of the $\rho$-meson.
Namely, even we use only very crude estimate for $g_1,g_2,g_3$, the magnetic and quadrupole form factors show very little dependence on these effective couplings near
$k^2\simeq 0$, so that a $500\,\%$ error in these couplings can be easily accommodated. Consequently,
checking these predictions in the lattice calculations would be
very instructive.

\item[vi)]
A priori, it is very hard to set an upper limit on the value of $k^2$
after which the NREFT formalism ceases to be applicable. An indirect check
a posteriori is however
possible. Namely, if the extracted values of $g_1(k^2),g_2(k^2),g_3(k^2)$ can be well approximated by a polynomial of a low order, then the formalism is likely to be valid in this range
of $k^2$.
\end{itemize}

\appendix

\begin{sloppypar}
{\em Acknowledgments:} The authors thank George Chanturia,
Jambul Gegelia, Christoph Hanhart, Maxim Mai, Alexey Nefediev and Kakha Shamanauri 
for interesting discussions, and José Antonio Oller for providing the parameters
of the IAM representation of the $\pi\pi$ amplitudes.
The work of A.S. and A.R. was funded in part by
the Ministry of Culture and Science of North Rhine-Westphalia through the NRW-FAIR project.
A.R. and U.-G.M. in addition, thank
the Chinese Academy of Sciences (CAS) President's
International Fellowship Initiative (PIFI) (grant nos. 2024VMB0001 and 2025PD0002) for the
partial financial support.
The work of J.-J.W. was supported by the
National Natural Science Foundation of China (NSFC) under Grants
No. 12221005, and by the Chinese Academy of Sciences under Grant No. YSBR-101.
Work partly funded by the Deutsche Forschungsgemeinschaft (DFG, German Research Foundation)
under Germany’s Excellence Strategy – EXC 3107 – Project-ID~533766364.
\end{sloppypar}

\section{Polarization vector of a spin-1 resonance}
\label{app:polarization}

In the case of a stable particle, the polarization vectors are well-defined.
A suitable basis is given by the three polarization vectors
$\varepsilon_\mu(P,s)$. In addition to orthogonality and completeness
relations, these vectors also satisfy the transversality condition:
\eq
\varepsilon^\mu(P,s) \varepsilon_\mu^*(P,s') = -\delta_{ss'}
&: \quad& \textrm{orthogonality}\, ,
\nonumber\\[2mm]
\sum_s \varepsilon_\mu(P,s) \varepsilon_\nu^*(P,s) =
-g_{\mu\nu} + \frac{P_\mu P_\nu}{P^2} &:\quad& \textrm{completeness},
\nonumber\\[2mm]
P^\mu \varepsilon_\mu(P,s) = 0 &:\quad& \textrm{transversality}.
\en
For a particle of mass $m$ with a nonzero momentum $\bm{P}=(0,0,P_3)$,
the polarization vectors in the Cartesian basis can be explicitly expressed as
\begin{equation}
\varepsilon^\mu(P,1) = 
\begin{pmatrix}
0 \\
1 \\
0 \\
0
\end{pmatrix}, \quad
\varepsilon^\mu(P,2) = 
\begin{pmatrix}
0 \\
0 \\
1 \\
0
\end{pmatrix}, \quad
\varepsilon^\mu(P,3) = \frac{1}{m}
\begin{pmatrix}
P_3 \\
0 \\
0 \\
\sqrt{\bm{P}^2 + m^2}
\end{pmatrix}\, ,
\end{equation}
whereas in the rotational basis they are equal to
\begin{equation}
\varepsilon^\mu(P,\pm 1) = \frac{1}{\sqrt{2}}
\begin{pmatrix}
0 \\
\mp 1 \\
-i \\
0
\end{pmatrix}, \quad
\varepsilon^\mu(P,0) = \frac{1}{m}
\begin{pmatrix}
P_3 \\
0 \\
0 \\
\sqrt{\bm{P}^2 + m^2}
\end{pmatrix}.
\end{equation}
It can be easily verified that these polarization vectors satisfy all the properties mentioned above.

In the case of an unstable particle, the situation is slightly more complex.
It is useful to think of $m^2\to s_R$ as a complex quantity, still keeping the components of the vector $\bm{P}$ real. The polarization vectors are then given by
\begin{equation}
\varepsilon^\mu(P,\pm 1) = \frac{1}{\sqrt{2}}
\begin{pmatrix}
0 \\
\mp 1 \\
-i \\
0
\end{pmatrix}, \quad
\varepsilon^\mu(P,0) = \frac{1}{\sqrt{s_R}}
\begin{pmatrix}
P_3 \\
0 \\
0 \\
\sqrt{\bm{P}^2 + s_R}
\end{pmatrix},
\end{equation}
and the conjugated vectors are defined as
\begin{equation}
\tilde\varepsilon^\mu(P,\pm 1) = \varepsilon^{\mu*}(P,\pm 1), \quad
\tilde\varepsilon^\mu(P,0) = \varepsilon^\mu(P,0),
\end{equation}
These polarization vectors satisfy the following properties:
\eq
\varepsilon^\mu(P,s) \tilde\varepsilon_\mu(P,s') = -\delta_{ss'} &:\quad& \textrm{orthogonality}\,,
\nonumber\\[2mm]
\sum_s \varepsilon_\mu(P,s) \tilde\varepsilon_\nu(P,s) = -g_{\mu\nu} + \frac{P_\mu P_\nu}{s_R} &:
\quad& \textrm{completeness},
\nonumber\\[2mm]
P^\mu \varepsilon_\mu(P,s) = P^\mu \tilde\varepsilon_\mu(P,s)=0 &:\quad& \textrm{transversality}.
\en

\section{Matching of the couplings $g_1,g_2,g_3$ to ChPT amplitude}
\label{app:ChPT}

Here we estimate the size of the couplings $g_1,g_2,g_3$
from matching to the relativistic amplitude calculated in $SU(2)$ Chiral Perturbation
Theory (ChPT).
We would like to stress that we consider a {\em perturbative} matching at threshold,
far away from the position of the $\rho$-meson. It is assumed that the
numerical values of
$g_1,g_2,g_3$ do not depend much on a particular kinematics used in the matching.
This assumption is far from being self-evident and needs further scrutiny. However,
this is the only available estimate at present and one has to stick to it.
Further work in this direction (ChPT with explicit vector mesons)
is in progress and will be reported in a separate
publication~\cite{Shamanauri}.

It can be checked that the tree-level amplitude does not give a contribution
to $g_1,g_2,g_3$ through matching. A full-fledged calculation of the five-point function
at next-to-leading order is quite a challenging exercise. Furthermore, it is not evident
whether this major effort will eventually pay off in view of the approximations already
made.
For this reason, we have decided to use a shortcut that allows one to easily get a crude,
order-of-magnitude estimates of these couplings. Namely, note that one of the
contributions
to $g_1,g_2,g_3$ comes from the tree diagrams containing $O(p^4)$ low-energy
constants (LECs) $l_i$. 
The part of the Lagrangian containing these LECs is given, e.g., in Ref.~\cite{Knecht:1997jw}
\eq
\mathscr{L}^{(4)} &=& \frac{l_1}{4} \langle D^\mu U^\dagger D_\mu U \rangle^2 +
\frac{l_2}{4} \langle D^\mu U^\dagger D^\nu U \rangle \langle D_\mu U^\dagger D_\nu U \rangle
\nonumber\\[2mm]
&+& \frac{l_3}{16} \langle \chi^\dagger U + U^\dagger \chi \rangle^2 + 
\frac{l_4}{4} \langle D^\mu U^\dagger D_\mu \chi + D^\mu \chi^\dagger D_\mu U \rangle \nonumber\\[2mm]
&+& l_5 \langle U QF^{\mu\nu} U^\dagger QF_{\mu\nu} \rangle + 
\frac{il_6}{2} \langle D^\mu U^\dagger QF_{\mu\nu} D^\nu U + D^\mu U QF_{\mu\nu} D^\nu U^\dagger \rangle \nonumber\\[2mm]
&-&\frac{l_7}{16} \langle \chi^\dagger U - U^\dagger \chi \rangle^2\, .
\en
Here, $U$ denotes the
pion field matrix, $\chi$ contains scalar and pseudoscalar sources and $F^{\mu\nu}$ is
the electromagnetic field strength tensor. Furthermore,
the covariant derivative is defined as $D_\mu X = \partial_\mu X - i e A_\mu [Q,X]$ and
$Q$ denotes the charge operator. For simplicity, external vector and axial-vector
sources (except
the electromagnetic field) are put to zero from the beginning.
The parameters $l_i$ are the LECs of the NLO ChPT Lagrangian. 

The scattering amplitude for the process
$\pi^+(q_1)+\pi^0(q_2)+A^\alpha(k)\to \pi^+(p_1)+\pi^0(p_2)$, which is given by
Eq.~(\ref{eq:matrixelement}), can be obtained in ChPT by using the above Lagrangian at tree level and 
is given by 
\eq
\mathscr{M}^\alpha = \frac{8el_1}{F_\pi^4}\, (p_2 \cdot q_2)(p_1 +q_1)^\alpha 
+ \frac{4el_2}{F_\pi^4}\, (p_2^\alpha (q_2 \cdot (p_1 + q_1)) + q_2^\alpha (p_2 \cdot (p_1 + q_1)))\, ,
\en
where $F_\pi$ denotes the pion decay constant. We would like to stress here once
more that this
is only a {\em part} of the amplitude -- namely, the one containing the LECs $l_i$.
Apart from this,
there are loop contributions, whose divergences cancel against the divergent parts of
$l_i$. We do
not calculate these contributions explicitly.

Next, note that the momenta $p_1,p_2,q_1,q_2$ in the above expression can be
expressed in terms of the
center-of-mass and relative momenta according to Eq.~(\ref{eq:CM-relative}).
From the
right-hand part of Eq.~(\ref{eq:matrixelement}) one may further conclude that only the part of the
amplitude that is linear in the relative momenta $p,q$ contributes to the matching
condition for
$g_1,g_2,g_3$. Retaining this part and dropping the rest, we arrive at
\eq\label{eq:ChPT-matrixelement}
\mathscr{M}^\alpha &=&\frac{2e(2l_1-l_2)}{F_\pi^4} \biggl( (p \cdot q) (P+Q)^\alpha 
+ p_\mu q_\nu (k^\mu g^{\nu\alpha} - k^\nu g^{\mu\alpha}) \biggr)\, .
\en
As said above, we perform the matching at threshold, where $P^2=Q^2=4M_\pi^2$, $k^2=p^2=q^2=0$. Comparing Eq.~(\ref{eq:ChPT-matrixelement})
with Eq.~(\ref{eq:matrixelement}) in this kinematics,
we see that the structure with $g_{\mu\nu}(P+Q)^\alpha$ does not carry the factor
$k^2$
and hence does not contribute to the matching condition.\footnote{More precisely,
in the $\rho$-meson form factor, the contribution coming from such
$k^2$-independent term should cancel against the $Z$-factor. This follows
from the fact that
the invariant form factor $G_1(k^2)$ is normalized to unity at $k^2=0$ owing to
the Ward identity and, thus,
cannot depend on the LECs $l_1,l_2$. Since we are interested solely in the form factor,
one can merely ignore this $k^2$-independent term.} Identifying the couplings
$g_1,g_2,g_3$ then gives
\eq\label{eq:ChPT-matching}
eg_1 f^2(0) = 0\, ,\quad\quad
eg_2 f^2(0) = \frac{2e(2l_1 - l_2)}{F_\pi^4}\,,\quad\quad
eg_3 f^2(0) = 0\, .
\en
To estimate the value of the coupling $g_2$, we can use the values of the
{\em finite parts} of the LECs $l_1$ and $l_2$ from ChPT, which are determined either
from experimental data or lattice QCD calculations.\footnote{The divergent part, of course, cancels against the loop contribution.}
In order to determine $f^2(0)$, one can invoke one of the known parameterizations
of the $\pi\pi$ amplitude, for example, the Schenk
parameterization~\cite{Schenk:1991xe}
\eq
\tan\delta^1_1(s) = \frac{(s - 4M_\pi^2)^{3/2}}{4\sqrt{s} M_\pi^2} \left( a^1_1 + \tilde{b}^1_1\left( \frac{s - 4M_\pi^2}{4M_\pi^2} \right) 
+ c^1_1 \left( \frac{s - 4M_\pi^2}{4M_\pi^2} \right)^2 \right) \left( \frac{4M_\pi^2 - s^1_1}{s - s^1_1} \right)\,,\quad\quad
\en
where $s$ is the Mandelstam variable and
$a^1_1 = 0.037$ is the parameter corresponding to the P-wave scattering length (note however a different sign convention).
Alternatively, one can use the parameterizations from
Refs.~\cite{Heuser:2024biq,Oller:1998hw}. Determining numerically
$a_1^1$ from these parameterizations, one arrives at
$a_1^1=0.039$~\cite{Heuser:2024biq} and $a_1^1=0.037$~\cite{Oller:1998hw}. Owing
to a minor difference in these values, we use $a_1^1=0.037$ in the following.

Next, using Eqs.~(\ref{eq:cot}) and (\ref{eq:ChPT-matching}), one gets
\eq\label{eq:matching-ChPT}
g_2 = -\frac{M_\pi^2(2l_1^r-l_2^r)}{24\pi F_\pi^4a_1^1}\, , \quad\quad
g_1=g_3 = 0\, ,
\en
where we have replaced $l_i$ by their finite parts.
Further, using the following values of the LECs $l_1^r = -6.0 \cdot 10^{-3}$ and $l_2^r = 2.7 \cdot 10^{-3}$ at the renormalization scale equal to the $\rho$-meson
mass that can be obtained from~\cite{Gasser:1983yg,Bijnens:1997vq}, we
estimate for the coupling $g_2$ as
$g_2\simeq1.4\,\mbox{GeV}^{-2}=2.8\cdot 10^{-2}M_\pi^{-2}$ (error bars are not considered
here, owing to a purely heuristic character of the above estimate).
As already said above, the full ChPT amplitude at $O(p^4)$ contains loop contributions as well, which we do not evaluate. Here, for a very crude estimate, we assume that these contributions are of the same order of magnitude as the contributions coming from the $O(p^4)$ LECs.

Finally, note that the couplings $g_1,g_3=0$ from matching to the ChPT amplitude at $O(p^4)$ (at least, to its polynomial part). This fact can be understood by noting that the kinematic factor in front of the form factor $G_2(k^2)$ contains one momentum, whereas the
other two structures depend on three momenta, see Eq.~(\ref{eq:scalar_ffs}). For this
reason, one may conclude that $g_1,g_3$ receive contributions starting from higher orders. This leads to the following order-of-magnitude estimate
\eq\label{eq:g_13}
|g_1|\simeq|g_3|\simeq\frac{|g_2|}{(4\pi F_\pi)^2}\, .
\en
Finally, we would like to stress once more that this is a very crude estimate. The chiral
logarithms, which are not taken into account, can easily lead to large factors in this
estimate. 

\section{Parameterization of the $\pi\pi$ amplitude in the vicinity of the pole}
\label{app:chanturia}

\begin{figure}[t]
\begin{center}
\includegraphics[width=8.cm]{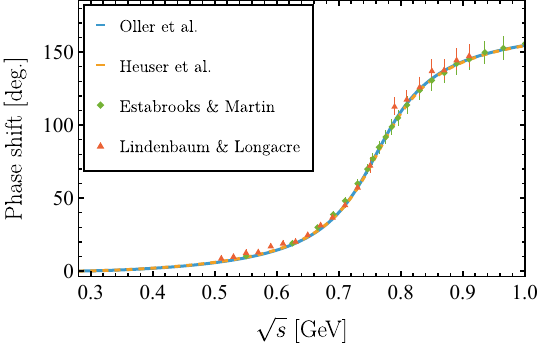}
\caption{The P-wave $\pi\pi$ phase shift in the channel with total isospin $I=1$,
the parameterizations from Ref.~\cite{Heuser:2024biq} and \cite{Oller:1998hw}.
Experimental data are taken from Refs.~\cite{Protopopescu:1973sh,Estabrooks:1974vu}.}
\label{fig:phaseshift}
\end{center}
\end{figure}

The Schenk parameterization~\cite{Schenk:1991xe}, which was used for the estimate
of the low-energy couplings, is not very convenient for the analytic continuation in
the complex plane. To this end, in order to evaluate the loop integral, we use
the ones taken from Refs.~\cite{Heuser:2024biq,Oller:1998hw}.
In particular, in Ref.~\cite{Heuser:2024biq}, the
scattering amplitude in the P-wave is proportional to
\eq
T(s)=\frac{1}{s-m^2+g^2\Sigma(s)}\, ,
\en
with the loop function given by
\eq
\Sigma(s)=\frac{s-4M^2}{48\pi^2}\,\int_{4M^2}^\infty
\frac{ds'\sqrt{s'-4M^2}}{\sqrt{s'}}\,\frac{\xi(s')}{s'-s}\, ,
\en
Here, $\xi(s')$ is a smooth cutoff that renders the integral UV finite
\eq
\xi(s')=\frac{1}{1+\dfrac{s'-4M^2}{s_B-4M^2}}\, ,\quad\quad s_B=(2\,\mbox{GeV})^2\, .
\en
The parameters $g$ and $m^2$ are chosen so that the above amplitude has a pole
on the second Riemann sheet exactly at
$s_R=\left(M_R-i{\Gamma_R}/{2}\right)^2$. Here, $M_R=0.7625\,\mbox{GeV}$ and
${\Gamma_R}/{2}=0.0732\,\mbox{GeV}$. This gives
\eq
g^2&=&-\frac{\mbox{Im}(s_R)}{\mbox{Im}(\Sigma(s_R))}\, ,
\nonumber\\[2mm]
m^2&=&\mbox{Re}(s_R)+g^2\mbox{Re}(\Sigma(s_R))\, .
\en
Note that $\Sigma(s_R)$ in the above expression should be evaluated on the second sheet.

The P-wave amplitude in the Ref.~\cite{Oller:1998hw} is given by
\eq
T=T_2(T_2-T_4^P-T_2GT_2)^{-1}T_2\, ,
\en
where all quantities are $2\times 2$ matrices describing the scattering
in $\pi\pi$ and $\bar KK$ coupled channels. The polynomial part of the amplitudes
$T_2$ and $T_4^P$ is given in Appendix B of Ref.~\cite{Oller:1998hw}, and
the free Green function $G$ is given in Eqs.~(32),(33) of the same paper.
The following values of the input masses and the coupling constant was used:
$M_\pi = 0.13957\,\mbox{GeV}$,
$M_K = 0.4957\,\mbox{GeV}$,
$F_\pi = 0.0931\,\mbox{GeV}$,
$F_K/F_\pi=1.22$. The values of the $O(p^4)$ coupling constants are given
in Table~II of Ref.~\cite{Oller:1998hw} and the cutoff momentum $q_{\sf max}=1.02\,\mbox{GeV}$. The pole in the amplitude
is located at $\sqrt{s_R}=(0.76080 - 0.07194i)\,\mbox{GeV}$.

For comparison, in Fig.~\ref{fig:phaseshift} we plot the phase shift obtained from
the two different parameterizations. It is seen that they agree very well, at least up to
$1\,\mbox{GeV}$.

\bibliographystyle{unsrt}
\bibliography{ref}

\end{document}